\documentclass[fleqn,usenatbib]{mnras}

\usepackage[varg]{newtxmath}

\usepackage[T1]{fontenc}
\usepackage{ae,aecompl}

\usepackage{graphicx}	
\usepackage{amsmath}	
\usepackage[dvipsnames]{xcolor}
\hypersetup{
  colorlinks   = true, 
  urlcolor     = RoyalBlue, 
  linkcolor    = RoyalBlue, 
  citecolor    = MidnightBlue 
}
\usepackage{natbib}
\usepackage{color}
\usepackage{rotating}
\usepackage{url}
\usepackage{ifthen}

\usepackage{bm}
\usepackage{bbold}
\usepackage{aas_macros}
\usepackage{verbatim}
\usepackage{algorithm}
\usepackage{algpseudocode}
\usepackage{enumitem}
\usepackage{subfig}

\usepackage{tablefootnote}
\usepackage[flushleft]{threeparttable}

\DeclareMathAlphabet{\mathpzc}{OT1}{pzc}{m}{it}

\newcommand\Dtilde{\stackrel{\sim}{\smash{\bm{\mathcal{D}}}\rule{0pt}{1.1ex}}}

\newcommand{\mvec}[1]{\bm{#1}}

\newcommand{\myvec}[1]{\mvec{#1}}

\SetSymbolFont{symbols}{bold}{OMS}{cmsy}{b}{n}
\DeclareSymbolFont{bmisymbols}{OML}{cmm}{b}{it}

\title[Simulation-based inference of cluster masses]{Simulation-based inference of dynamical galaxy cluster masses with 3D convolutional neural networks}

\author[D. Kodi Ramanah, R. Wojtak, N. Arendse]{Doogesh Kodi Ramanah,$^{1}$\thanks{ramanah@nbi.ku.dk} Rados\l{}aw Wojtak,$^{1}$\thanks{radek.wojtak@nbi.ku.dk} Nikki Arendse$^{1}$\thanks{nikki.arendse@nbi.ku.dk}\\
$^{1}$ DARK, Niels Bohr Institute, University of Copenhagen, Jagtvej 128, 2200 Copenhagen, Denmark
}

\date{Accepted XXX. Received YYY; in original form ZZZ}

\pubyear{2020}

\begin{document}
\label{firstpage}
\pagerange{\pageref{firstpage}--\pageref{lastpage}}
\maketitle

\begin{abstract}
We present a simulation-based inference framework using a convolutional neural network to infer dynamical masses of galaxy clusters from their observed 3D projected phase-space distribution, which consists of the projected galaxy positions in the sky and their line-of-sight velocities. By formulating the mass estimation problem within this simulation-based inference framework, we are able to quantify the uncertainties on the inferred masses in a straightforward and robust way. We generate a realistic mock catalogue emulating the Sloan Digital Sky Survey (SDSS) Legacy spectroscopic observations (the main galaxy sample) for redshifts $z \lesssim 0.09$ and explicitly illustrate the challenges posed by interloper (non-member) galaxies for cluster mass estimation from actual observations. Our approach constitutes the first optimal machine learning-based exploitation of the information content of the full 3D projected phase-space distribution, including both the virialized and infall cluster regions, for the inference of dynamical cluster masses. We also present, for the first time, the application of a simulation-based inference machinery to obtain dynamical masses of around $800$ galaxy clusters found in the SDSS Legacy Survey, and show that the resulting mass estimates are consistent with mass measurements from the literature.
\end{abstract}




\section{Introduction}
\label{intro}

Galaxy clusters are formed by the collapse of high density regions in the early Universe, and they are important to study the formation and evolution of large-scale cosmic structures. The cluster abundance as a function of mass and its evolution are sensitive to the amplitude of density perturbations and to the properties of dark matter and dark energy. Galaxy clusters can therefore provide competitive cosmological constraints that are complementary to other cosmological probes. As future surveys, such as the Dark Energy Spectroscopic Instrument \citep[DESI,][]{DESI2016}, the Vera C. Rubin Observatory \citep{lsst2008summary}, Euclid \citep{euclid2016missiondesign} and eROSITA \citep{eROSITA2012}, will provide unprecedented volumes of data extending to high redshifts, the accuracy and precision of cluster mass estimation techniques will become crucial. With the ever increasing scale of state-of-the-art cosmological simulations \citep[e.g.][]{paco2019quijote, uchuu2020DR1} providing considerable volumes of training data, along with the limitations of traditional techniques, the use of machine learning (ML) algorithms to infer cluster masses has become an increasingly attractive and viaile option \citep[e.g.][]{sutherland2012SDM, ntampaka2015machine, ntampaka2016dynamical, armitage2019application, calderon2019prediction, ho2019robust, ho2020approximate, DKR2020NF, yan2020galaxy}. These models are typically trained on a large simulated data set, such that the algorithm learns the connection between the observables and cluster masses. Once optimized, they can subsequently be used to predict masses for unseen data, provided that the simulations used for training are sufficiently accurate to replicate the characteristics of the galaxy survey of interest with high fidelity \citep{cohn2020multiwavelength}.

\medskip
For observations probing galaxy kinematics in galaxy clusters, ML methods offer a promising alternative to traditional methods of cluster mass estimation which are usually based on scaling relations, the virial theorem or the Jeans equation, and are limited by several assumptions, primarily involving dynamical equilibrium and spherical symmetry, as briefly reviewed in \citet{DKR2020NF}. Recently, convolutional neural networks (CNNs), by virtue of their sensitivity to visual features, have been applied by \citet{ho2019robust} to obtain accurate dynamical mass estimates of galaxy clusters in spectroscopic surveys. The network inputs are images generated by a kernel density estimator from the 2D projected phase-space distributions defined by the cluster-centric projected distance and line-of-sight velocities of galaxies observed in the fields of clusters. The challenge with ML methods is often to not only produce single point estimates, but also a reliable estimate of the associated uncertainties. The most recent attempts to approach the problem of uncertainty estimation used normalizing flows \citep{DKR2020NF} to infer the conditional probability distribution of the dynamical cluster masses and approximate Bayesian inference to assign prior distributions to the neural network weights \citep{ho2020approximate}. Despite the increasing popularity of ML-based methods, the classical techniques of cluster mass estimation still currently prevail over the applications to observational data. Nevertheless, they are likely to be superseded by their ML counterparts for future applications involving next-generation surveys.

\medskip
The primary challenge intrinsic to galaxy cluster mass estimation is posed by interlopers. These are galaxies that are not gravitationally bound to the cluster, but that are located along the line of sight and have similar line-of-sight velocities to the cluster. Distinguishing interlopers from member galaxies is a problematic task, because redshift surveys can only provide information about the positions and velocities of objects along the line of sight, and not perpendicular to it. Finding an effective way of reducing contamination from interlopers, with the limited information available from surveys, is essential to improve the accuracy of galaxy cluster mass estimates.

\medskip
In this work, we propose to work at the level of 3D projected phase-space distribution, characterized by the sky projected galaxy positions and their line-of-sight velocities, instead of the standard 2D phase-space, to alleviate the interloper contamination and improve the precision of cluster mass estimation, as motivated by the following arguments. Cluster members are distributed more symmetrically around the cluster centre, while interlopers can clump in any place. Moreover, 2D phase-space density is averaged over the position angle, such that the information on any axially asymmetric localization of interlopers is lost, rendering it more difficult for the algorithm to differentiate between interlopers and cluster members. In contrast, 3D phase-space density retrieves the information encoded in the position angle and is, therefore, expected to provide a better separation between cluster members and interlopers. Moreover, dynamical substructures have been shown to result in an artificial overestimation of cluster masses \citep{tucker2018galaxy,old2018clusterIII}. These substructures may also induce an asymmetry within the boundary of dark matter halos, such that the 3D phase-space density will more adequately account for the presence of substructures and help to mitigate this bias. To optimize the information from the 3D dynamical phase-space distribution, we make use of 3D convolutional kernels, naturally designed to extract spatial features, in neural networks. Compared to the previous studies, we also adopt larger apertures than the virial sphere. This allows us to include the cluster infall zone as an extra constraining power in the estimation of dynamical masses. The observed infall patterns around galaxy clusters have long been used to measure cluster mass profiles at large distances \citep{Diaferio1997,Diaferio1999,Rines2003,Falco2014}.

\medskip
We opt for a simulation-based inference approach to quantify the uncertainties on the neural network predictions. Simulation-based inference \citep[e.g.][and references therein]{cranmer2019frontier}, often referred to as {\it likelihood-free inference}, encompasses a class of statistical inference methods where simulations are used to estimate the posterior distributions of the parameters of interest conditional on data, without any prior knowledge or assumption of the likelihood distribution. Simulation-based inference has emerged as a viable alternative to perform Bayesian inference under complex generative physical models using only simulations. This framework allows all physical effects encoded in forward simulations to be properly accounted for in the inference pipeline, without having recourse to inadequate or misguided likelihood assumptions. As such, simulation-based inference, and variants thereof, have recently garnered significant interest for cosmological data analysis \citep[e.g.][]{akeret2015ABC, lintusaari2017elfi, jennings2017astroABC, leclercq2018bolfi, charnock2018IMNN, alsing2018massive, alsing2019fast, alsing2019nuisance, Wang2020}.

\medskip
In essence, we present a simulation-based inference framework for the estimation of the dynamical mass of galaxy clusters with 3D convolutional neural networks. The approach presented here is complementary to our previous neural flow (NF) mass estimator \citep[][hereafter NF2020]{DKR2020NF} in various aspects. This is primarily a conceptually different framework of uncertainty estimation using neural networks. The simulation-based inference machinery, as presented here, allows the inference of the approximate posterior distribution of cluster masses given their 3D projected phase-space distribution, using an ensemble of simulated clusters and a neural network designed to extract summary statistics. In contrast, the NF mass estimator is a neural density estimator, where the cluster mass inference problem is formulated within a conditional density estimation framework. The two methods also differ in network architecture and dimensionality of their respective inputs. The approach presented here employs 3D convolutional kernels to fully exploit the information encoded in the 3D phase-space distribution of galaxy clusters, while the NF mass estimator relies on fully connected layers, i.e. multilayer perceptrons, and works at the level of the compressed 2D phase-space dynamics.

\medskip
The remainder of this manuscript is organized as follows. Section~\ref{dynamical_3D_phase_space_distribution} provides an overview of the 3D dynamical phase-space distribution in terms of the key observables used for training the neural network. We also outline the mock generation procedure for cluster catalogues emulating the features of the actual SDSS data set and the preprocessing steps involved in the preparation of the training and test sets. We then describe the simulation-based inference approach utilized in this work, followed by a brief introduction to convolutional neural networks and a description of the neural network architecture and training procedure in Section~\ref{SBI_CNN}. We subsequently validate and demonstrate the performance of the optimized model on the test cluster catalogue in Section~\ref{validation_performance} and follow up by inferring cluster masses from the actual SDSS catalogue in Section~\ref{applications_sdss}. Finally, we provide a summary of the main aspects and findings of our work in Section~\ref{conclusions}, and highlight potential future investigations with cosmological applications.

\section{Dynamical phase-space distribution}
\label{dynamical_3D_phase_space_distribution}

We outline the general problem of cluster mass estimation by first introducing the dynamical phase-space distribution. We then describe the generation of the mock SDSS catalogue which will be used to train and evaluate the performance of the neural network in future sections.

\subsection{Galaxy cluster observables}
\label{galaxy_cluster_observables}

The definition of cluster's halo mass adopted throughout this work is $M_{200c}$, corresponding to the mass contained in a sphere with mean density equal to 200 times the critical density of the Universe at the halo's redshift. We obtain an estimate of the mass by employing the full projected phase-space distribution of galaxy clusters. This consists of the positions of each member galaxy projected onto the ($x, y$) plane of the sky, denoted as ($x_{\rm{proj}}, y_{\rm{proj}}$), as well as their separate line-of-sight velocities, $v_{\rm{los}}$, as provided by redshift surveys. In this work, instead of computing the projected radial distance from the cluster centre as $R_{\rm{proj}} = (x_{\rm{proj}}^2 \, +\,  y_{\rm{proj}}^2)^{1/2}$ as is typically done, we exploit the information from $x_{\rm{proj}}$ and $y_{\rm{proj}}$ separately. This should make our model more sensitive to interlopers and substructures, which are often located asymmetrically around the cluster centre. We adopt units of $h^{-1} \rm{Mpc}$ for $x_{\rm{proj}}$ and $y_{\rm{proj}}$ throughout this work.

\subsection{Mock cluster catalogues}
\label{mock_catalogues}

\begin{figure}
	\centering
		{\includegraphics[width=\hsize,clip=true]{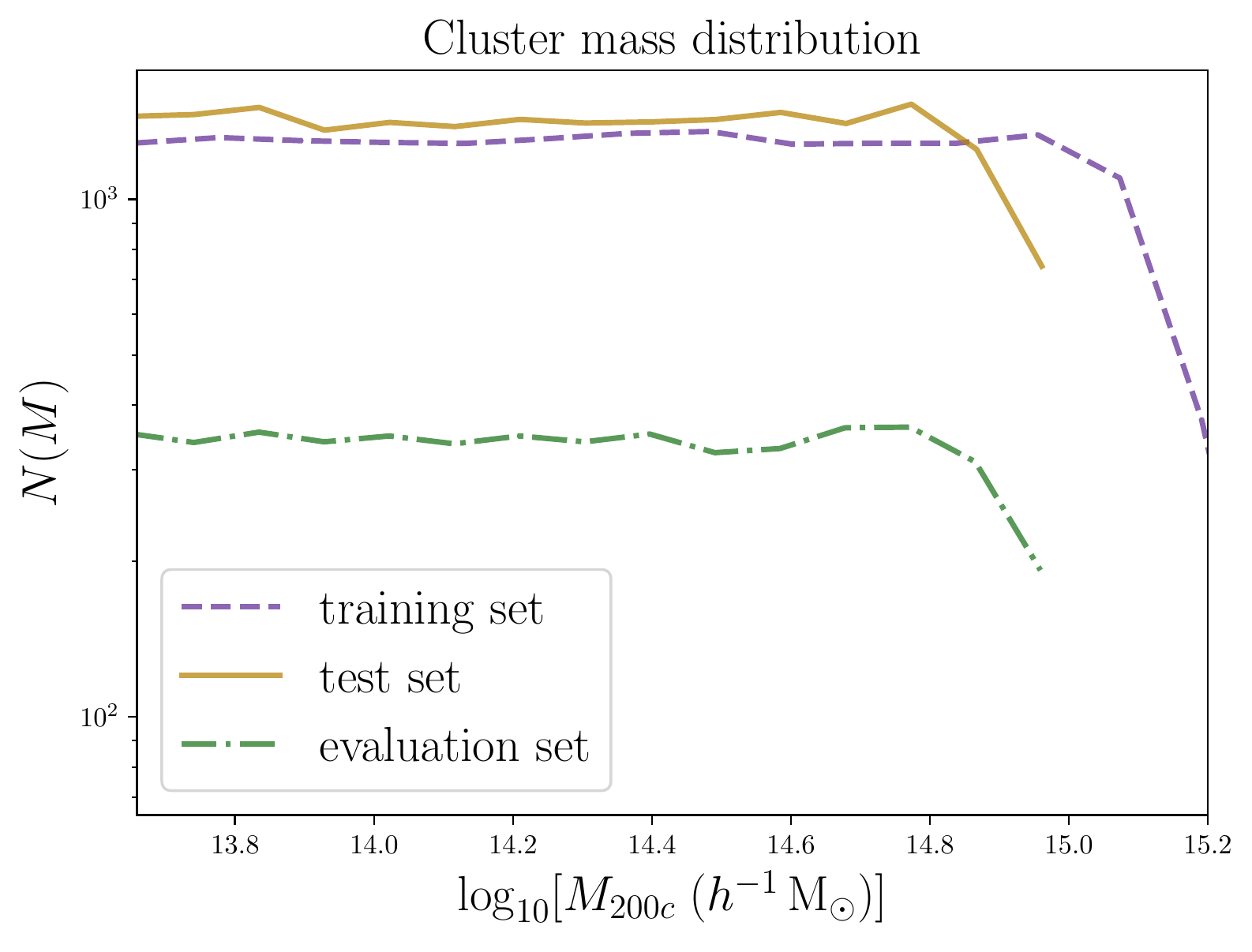}}
	\caption{Cluster mass distribution, i.e. variation of number of clusters with logarithmic mass for the training, test and evaluation sets extracted from the mock SDSS catalogue. To ensure we do not induce any cosmological information or selection bias while training the neural network, we upsample the relatively scarce high-mass clusters using independent lines of sight, thereby resulting in an approximately flat mass distribution for the training set.}
	\label{fig:cluster_mass_function}
\end{figure}

We generate mock observations of galaxy clusters using publicly available galaxy catalogues derived from the \textsc{MultiDark} simulations \citep{klypin2016multidark}.\footnote{http://skiesanduniverses.org} Among the three different semi-analytic galaxy formation models applied to the simulation \citep{Knebe2018}, we opted for Semi-Analytic Galaxies (\textsc{sag}), which includes the most complete implementation of modelling orphan galaxies and, therefore, produces the most realistic distribution of galaxies in the cluster cores \citep{Cora2006, cora2018sag}. For more details regarding implementations of the star formation and feedback processes in \textsc{sag} as well as a comparison to the remaining two semi-analytic models, i.e. \textsc{galacticus} \citep{Benson2012} and the Semi-Analytic Galaxy Evolution (\textsc{sage}) model \citep{Croton2006}, we refer the interested reader to \citet{Knebe2018}. The galaxy catalogues from \textsc{sag} contain the positions and absolute magnitudes in the SDSS filters at all snapshots of the simulation. The background dark matter simulation (MDPL2) was run for the Planck $\Lambda$CDM cosmological model \citep{planck2014cosmo}. The simulation box has a size of $1 \, h^{-1}$~Gpc and a mass resolution of $1.51\times10^9 \, h^{-1} \, {\rm M}_{\odot}$.

\medskip
We select galaxy clusters as massive dark matter halos found in the halo catalogues produced by the \textsc{rockstar} halo finder \citep{behroozi2013rockstar}. For every halo, we construct its cluster's projected phase-space diagram by drawing a line of sight and computing the corresponding projections of the galaxy positions and velocities onto the plane of the sky and the line of sight, respectively. All phase-space coordinates are calculated relative to the central galaxy assigned to the main cluster halo and the observed velocities include the Hubble flow with respect to the cluster centre. The final projected phase-space diagrams are generated by applying the following cuts: $\pm 2200$~km~s$^{-1}$ in line-of-sight velocities $v_{\rm los}$ and $\pm 4\, h^{-1} \rm{Mpc}$ in proper distances $x_{\rm proj}$ and $y_{\rm proj}$.

\medskip
Aiming at generating mock data which resemble the main spectroscopic galaxy sample of the SDSS Legacy Survey \citep{Strauss2002}, we adopt a flux limit of 18.0 magnitude in $r$-band. The flux limit is 0.2 magnitude lower than the actual SDSS limiting magnitude in order to compensate the slightly lower counts of galaxies in simulated clusters than in the SDSS ones \citep[see][]{Knebe2018}. The apparent magnitudes of all galaxies in the field of each galaxy cluster are computed by assigning each simulated cluster an observer located at comoving distance randomly drawn from a uniform distribution within a 3D ball. The maximum comoving distance is $250 \, h^{-1}$~Mpc, for which galaxy cluster detection in the SDSS main galaxy sample is complete down to a cluster mass of $\sim 10^{14.0}h^{-1}{\rm M}_{\odot}$ \citep{abdullah2020cosmo}.

\medskip
Our mock SDSS galaxy catalogue is generated assuming completeness of spectroscopic observations down to the assumed flux limit. This is an idealized assumption because the actual completeness of the SDSS decreases in high-density regions due to the physical limit on the minimum distance between SDSS fibres. However, since our CNN mass estimator operates on smoothed KDE density maps, we expect that downsampling due to incompleteness of SDSS spectroscopic observations should not have a noticeable impact on the final mass estimates. The insensitivity of CNN mass estimators based on smoothed density maps to stochastic downsampling was shown in \citet{ho2019robust} and \citet{DKR2020NF}. This test can be repeated for a density-dependent incompleteness resembling the SDSS selection for spectroscopic observations. Considering an extreme case when cluster data are missing spectroscopic velocities inside cluster cores subtending 55 arc secs, which is the minimum distance between SDSS fibres \citep{Strauss2002}, we find that, for a sample of SDSS-like clusters, our CNN trained on complete mocks yields mass estimates only $0.017$~dex lower in average. Since the SDSS is more complete than this extreme example, primarily by virtue of an optimized tiling, we conclude that a realistic bias is even smaller and currently negligible compared to the precision of our mass estimator.

\medskip
Keeping in mind possible future applications of our dynamical mass estimator for cosmological inference with the cluster abundance, it is instructive to generate a galaxy cluster sample for which the distribution of cluster masses is independent of cosmological model through the mass function. An optimal solution is to consider a set of clusters with a flat distribution in log mass within a possibly wide range of dynamical cluster masses. Aiming at generating a sample with $\sim 10^{4}$ galaxy clusters, we downsample the actual mass function below halo mass $M_{\rm 200c} \approx 10^{14.3} h^{-1} {\rm M}_{\odot}$ and generate up to 25 projections per cluster at higher masses. In order to minimize correlations between projected phase-space diagrams derived from the same cluster, we use a set of directions (up to 25 lines of sight) found by maximizing angular separations between every two closest sight lines. The adopted maximum number of sight lines per cluster is not sufficient to keep a flat distribution at the high-mass end, i.e. $\log_{10}M_{200c}\gtrsim14.9$ (cf. Fig.~\ref{fig:cluster_mass_function}). This, however, can hardly be improved because further increase of upsampling would introduce strong correlations between phase-space diagrams generated from the same galaxy cluster. The final sample contains $4.3\times 10^{4}$ galaxy clusters with a minimum halo mass of $10^{13.7}{h^{-1} \, \rm M}_{\odot}$.

\medskip
The overdensity threshold used in the halo mass definition depends on redshift. This leads to a well-known non-physical evolution of halo masses which reflects merely the redshift dependence of the critical density \citep{Diemer2013}. Since phase-space diagrams do not provide any information on cluster redshifts required to adjust the overdensity threshold, mass estimates from neural networks may be consequently affected by an additional noise. For a wide redshift range, the noise may be sufficiently large so that it would be necessary to supplement each phase-space diagram with the information on cluster redshift setting the corresponding overdensity threshold. However, for our mock data spanning a relatively narrow redshift range $z\leq 0.085$, the expected uncertainty due to the lack of information on cluster redshifts amounts to only 0.006~dex, which is significantly lower than the level of precision obtained in our work and similar studies, i.e. $\sim 0.1$~dex.

\medskip
We extract the training set, with a flat mass distribution, containing around seventeen thousand clusters by randomly drawing from the mock catalogue. The corresponding validation set, used for early stopping when optimizing the neural network, is designated as $10\%$ of the training set, such that it contains $\sim 1700$ clusters and only $\sim 15500$ clusters are utilized during training. The test set consisting of twenty thousand clusters is obtained by randomly sampling from the remaining clusters in the mock catalogue. The remaining $\sim 5000$ clusters in the catalogue then constitute an evaluation set. The test set is used in the simulation-based inference framework (cf. Section~\ref{SBI_sub}), whilst the purpose of the evaluation set is to assess the performance of the network (cf. Section~\ref{neural_network_evaluation}). The mass distributions of the non-overlapping training, test and evaluation sets are depicted in Fig.~\ref{fig:cluster_mass_function}.

\subsection{Kernel density estimator}
\label{kde}

\begin{figure}
	\centering
    \subfloat{\includegraphics[width=0.90\hsize]{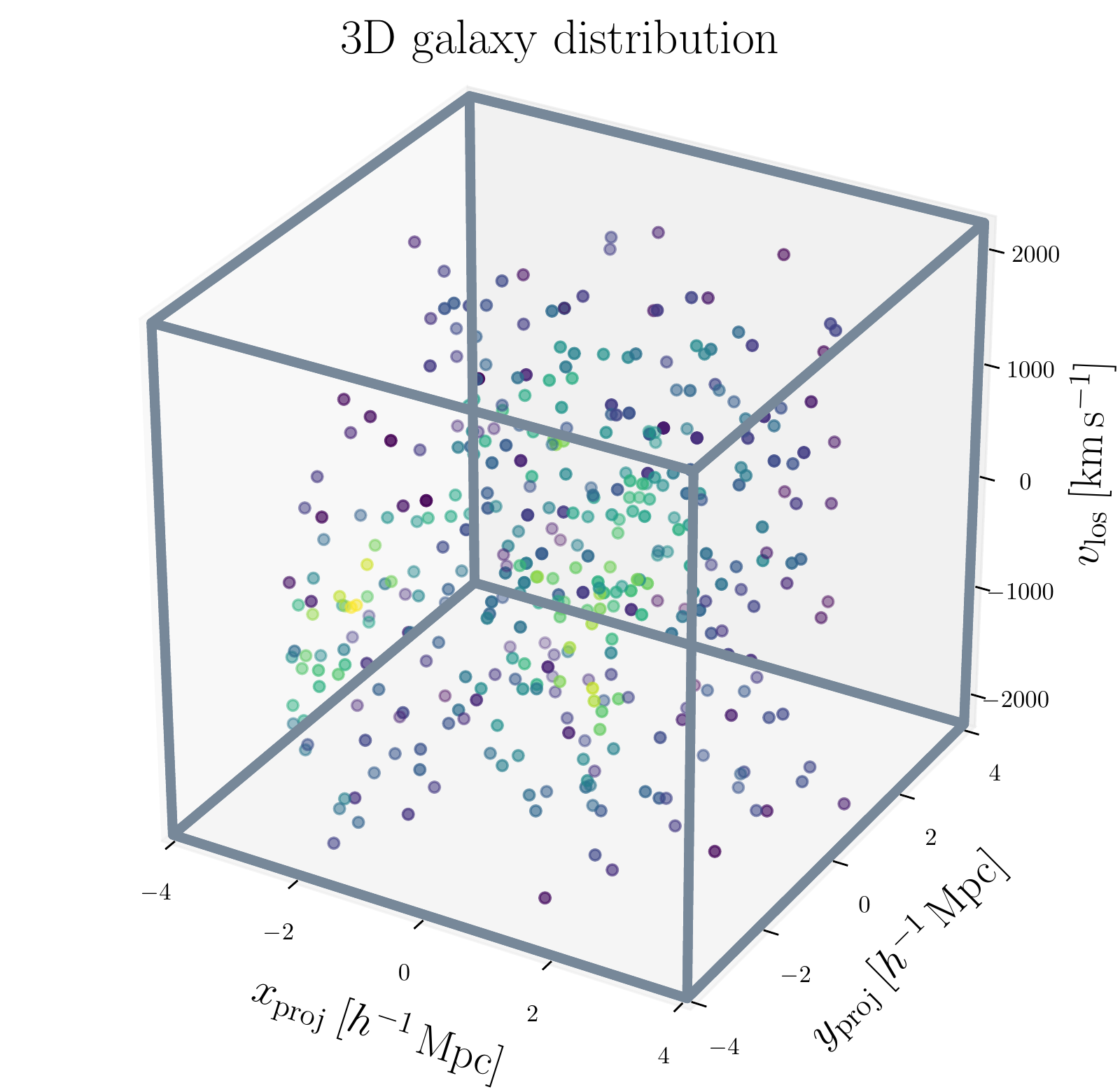}}
    \qquad \qquad \qquad 
    \subfloat{\includegraphics[width=0.72\hsize]{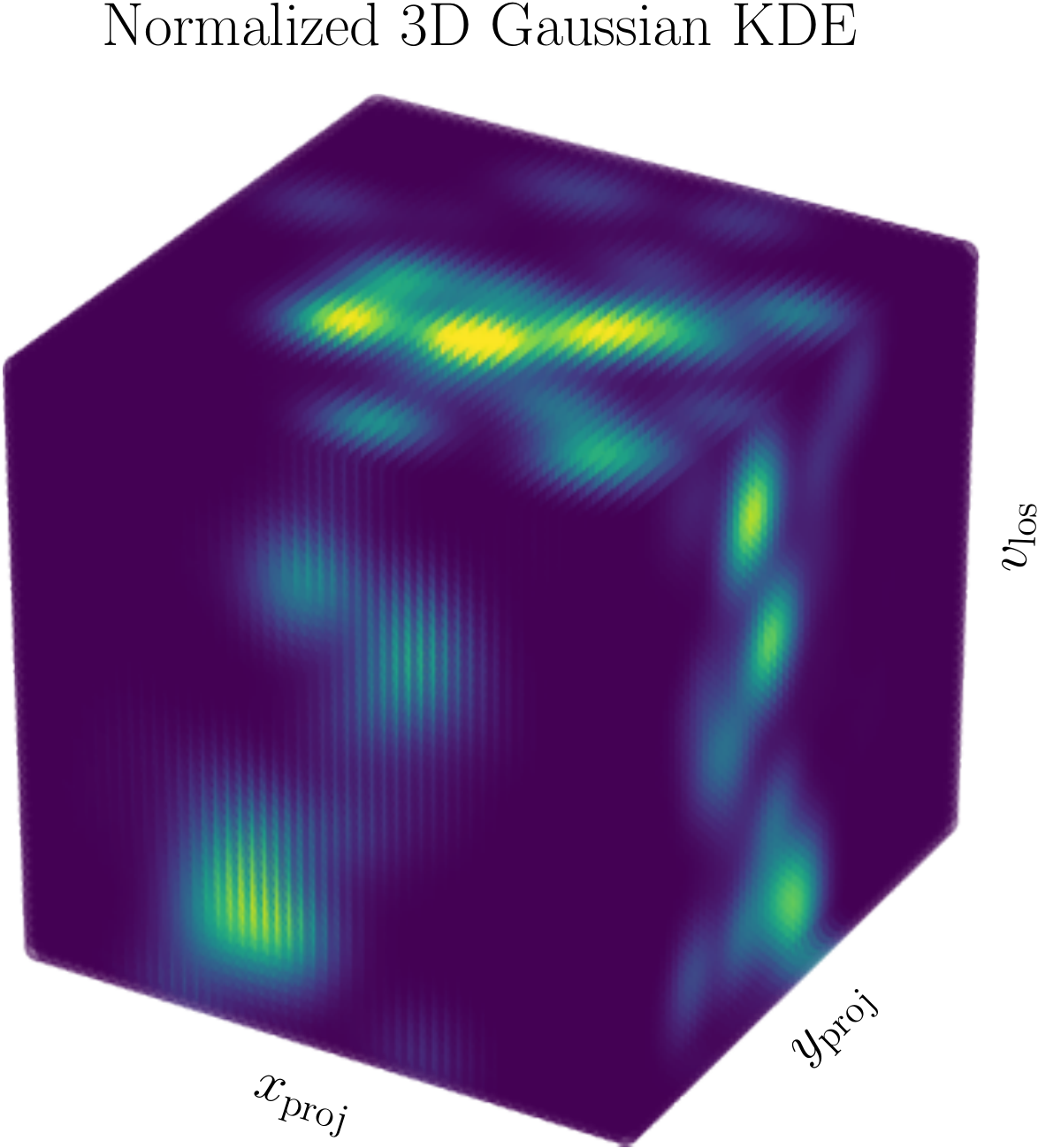}}
    \quad
	\caption{Simulated 3D phase-space distribution of galaxies observed in an example cluster of galaxies, represented via its 3D galaxy distribution ({\it top panel}) and 3D Gaussian KDE ({\it bottom panel}), consisting of projected galaxy positions in the sky, $x_{\mathrm{proj}}$ and $y_{\mathrm{proj}}$, and line-of-sight velocities, $v_{\mathrm{los}}$, in dynamical phase space. The KDE representation serves as inputs to our 3D convolutional neural network.}
	\label{fig:phase_space_distribution}
\end{figure}

Before the observables ($v_{\rm{los}}, \, x_{\rm{proj}}, \, y_{\rm{proj}}$) are provided as input to the ML model, they are first preprocessed with a kernel density estimator (KDE) to create a smooth PDF mapping in the 3D phase-space distribution. This is done in order to obtain similar-sized arrays as inputs for all clusters, which contain different numbers of member galaxies, as well as to create a visual input (image) for the convolutional neural network. An in-depth review of kernel density estimation is provided in \citet{diggle1984monte, wand1994kernel, sheather2004density}.

\medskip
Let our complete set of $n$ observables $\{ \myvec{X}_1, \myvec{X}_2, \ldots, \myvec{X}_n \}$, in which each variable $\myvec{X}_i$ is given by ($v_{\rm{los}}, \, x_{\rm{proj}}, \, y_{\rm{proj}}$), be drawn from an unknown distribution with density $f$. The density evaluated at a point $\myvec{x} = (v_{\rm{los}}, \, x_{\rm{proj}}, \, y_{\rm{proj}})$ can be approximated as
\begin{equation}
\hat{f}(\myvec{x}) = \frac{1}{n | \mathbf{H} |^{1/2} } \sum^n_{i=1} \, K \big[ \mathbf{H}^{-1/2} (\myvec{x} - \myvec{X}_i) \big], \label{eq:KDE_definition}
\end{equation}
where $K$ is the kernel function and $\mathbf{H}$ is a $3\times3$ bandwidth matrix. The KDE sums up the density contributions from the collection of data points $\{ \myvec{X}_1, \myvec{X}_2, \ldots, \myvec{X}_n \}$ at the evaluation point $\myvec{x}$. Data points close to $\myvec{x}$ contribute significantly to the total density, while data points further away from $\myvec{x}$ have only a relatively small contribution. The shape of the density contributions is determined by the kernel function, and their size and orientation are dictated by the bandwidth matrix. In this work, we use a 3-dimensional Gaussian kernel given by
\begin{equation}
K(\myvec{u}) = (2 \pi)^{-3/2} \, |\mathbf{H}|^{-1/2} \; \exp \left(- \tfrac{1}{2} \, \myvec{u}^\intercal \, \mathbf{H}^{-1} \, \myvec{u} \right), \label{eq:kernel_definition}
\end{equation}
with $\myvec{u} = \myvec{x} - \myvec{X}_i$. For the bandwidth matrix, a scaling factor $\kappa$ is multiplied with the covariance matrix of the data, $\mathbf{H} = \kappa \mathbf{\Sigma}$. The scaling factor should be sufficiently small to encapsulate even the more subtle features of the data and small-scale signal expected for low-mass clusters, but large enough that the ML model is robust to changes in galaxy number count, and can easily interpolate between the data sets of discrete and quite scarcely distributed points. We performed some numerical experiments with three distinct scaling factors, $\kappa = \{ 0.15, 0.175, 0.20 \}$, and found only a marginal influence on the model predictions. We, therefore, opted for the intermediate value of $\kappa = 0.175$ in our study.

\medskip
An example of a 3D KDE representation that serves as input to the neural network for one particular cluster is illustrated in Fig.~\ref{fig:phase_space_distribution}. For all clusters considered in this work, unless otherwise stated, the extents of the observables are as follows: $v_{\mathrm{los}} \in [-2200, 2200] \, \rm{km} \, \rm{s}^{-1}$, $x_{\mathrm{proj}} \in [-4.0, 4.0] \, h^{-1} \rm{Mpc}$ and $y_{\mathrm{proj}} \in [-4.0, 4.0] \, h^{-1} \rm{Mpc}$, with 50 voxels along each axis, resulting in 3D slices of dimension $50^3$. Concerning the choice of the maximum extent for $x_{\mathrm{proj}}$ and $y_{\mathrm{proj}}$, we performed an optimization procedure for different sizes of $\{ 1.6, 4.0, 6.0 \} \, h^{-1} \rm{Mpc}$ and opted for $4.0 \, h^{-1} \rm{Mpc}$, which resulted in a noticeable improvement in precision of our mass estimator relative to a size of $1.6 \, h^{-1} \rm{Mpc}$ and virtually no loss of constraining power relative to a size of $6.0 \, h^{-1} \rm{Mpc}$.

\section{Simulation-based inference with neural networks}
\label{SBI_CNN}

In this section, we present the rationale underlying simulation-based inference and convolutional neural networks, and outline how a standard neural network may be employed within the simulation-based inference framework to robustly quantify the uncertainties on the model predictions. We also describe the implementation of our 3D convolutional neural network in terms of the network architecture and training routine.

\subsection{Simulation-based inference}
\label{SBI_sub}

Simulation-based inference (hereafter SBI) encompasses a class of statistical inference techniques employing a simulator, which inherently defines a statistical model, capable of generating high-fidelity simulations for comparison with actual observations \citep[see, for e.g.,][for an in-depth review of recent developments]{cranmer2019frontier}. However, the probability density for a given observation, i.e. the likelihood, which constitutes a crucial component of any statistical inference framework, is generally intractable, especially for problems involving complex dynamics as relevant to this work. SBI techniques, as relevant to this work, therefore, typically involve an estimation of the likelihood or posterior via informative summary statistics using classical density estimators \citep{diggle1984monte} or neural density estimators \citep[e.g.][]{rezende2015normalizing, germain2015made, uria2016neural, kingma2016IAF, papamakarios2016fast, papamakarios2017MAF, papamakarios2018SNL, huang2018NAF} from recent advances in ML. In essence, these density estimators are used to approximate the distribution of summary statistics of the samples generated from the simulator.

\medskip
The SBI framework adopted in this work is inspired by the approach presented in \citet{charnock2018IMNN}, where they demonstrate that parameter inference is feasible via SBI using summary statistics provided by a neural network. They developed an information maximizing neural network to produce optimal summary statistics, but the approach presented therein may be employed with any neural network predicted summaries. This is because a neural network, by design, performs some form of data compression (or dimensionality reduction) to extract meaningful features from a given input data set to yield informative summaries of the data. Formally, a neural network, $\mathbb{NN}(\myvec{\theta}, \myvec{\gamma}): \myvec{d} \rightarrow \tilde{\myvec{\tau}}$, may be described as a trainable and flexible approximation of a model, $\mathcal{M}: \myvec{d} \rightarrow \myvec{\tau}$. The neural network maps some input data $\myvec{d}$ to a prediction or estimate $\tilde{\myvec{\tau}}$ of the desired label or target $\myvec{\tau}$ associated with the data. It is parameterized by a set of trainable weights $\myvec{\theta}$ and a set of hyperparameters $\myvec{\gamma}$, which encompasses the choice of network architecture, initialization of the weights, type of activation and loss functions.

\medskip
During training, the network weights $\myvec{\theta}$ are optimized via stochastic gradient descent to minimize a particular cost or loss function given a training data set. The loss function is equivalent to the negative logarithm of the likelihood, $- \ln \mathcal{L} (\tilde{\myvec{\tau}} | \myvec{d}, \hat{\myvec{\theta}}, \hat{\myvec{\gamma}})$, for a given set of network weights and hyperparameters, $\myvec{\theta} = \hat{\myvec{\theta}}$ and $\myvec{\gamma} = \hat{\myvec{\gamma}}$, respectively. As such, the loss function provides a measure of how close the network prediction $\tilde{\myvec{\tau}}$ is to the desired target $\myvec{\tau}$. Training the neural network entails finding the maximum likelihood estimates $\hat{\myvec{\theta}}_{\mathrm{MLE}}$ of the network weights with a given training set of data-target pairs $\{ \myvec{d}, \myvec{\tau} \}$ at fixed hyperparameters $\hat{\myvec{\gamma}}$. In the ideal scenario, there would be one global minimum in the likelihood surface, but this is generally not the case in practice, with the surface being extremely complex, degenerate and non-convex. Consequently, it is highly probable that the weights will only converge to a local minimum on the likelihood surface, which is dictated to some extent by the initialized values. This is a well-known caveat inherent to the standard training routine for neural networks. SBI provides a means to mitigate this crucial problem with classically trained networks to obtain scientifically rigorous predictions, including reliable uncertainties, of the true targets $\myvec{\tau}$ given the input data $\myvec{d}$.

\medskip
For the particular mass inference problem studied here, the SBI approach entails the generation of an ensemble of galaxy clusters using a physical model or simulator. We employ some physical priors, in terms of a flat distribution in dynamical mass (as motivated by decoupling from the cosmological model imprinted in the halo mass function) and uniform spatial distribution, in the cluster generation procedure (cf. Section~\ref{mock_catalogues}). Given that our ultimate objective is the inference of cluster masses from the SDSS catalogue, we generate a realistic set of SDSS-like clusters. By feeding the 3D phase-space distributions of this set of generated clusters to our trained neural network, $\mathbb{NN}(\hat{\myvec{\theta}}, \hat{\myvec{\gamma}}): \myvec{d} \rightarrow \tilde{d}$, we obtain a corresponding set of predicted summaries $\tilde{d}$, which, by design, correspond to the cluster masses. This allows us to characterize the joint probability distribution of data (via the compressed summaries) and parameters, $\mathcal{P}( \tilde{d} , M )$, via a kernel (or neural) density estimator. By slicing this joint distribution at any observed data fed to the network, $\mathbb{NN}(\hat{\myvec{\theta}}, \hat{\myvec{\gamma}}): \myvec{d}_{\mathrm{obs}} \rightarrow \tilde{d}_{\mathrm{obs}}$, we obtain the approximate posterior as follows:
\begin{equation}
	 \mathcal{P}(M | \tilde{d}_{\mathrm{obs}}) \approx \mathcal{P}( M | \myvec{d}_{\mathrm{obs}}, \hat{\myvec{\theta}}, \hat{\myvec{\gamma}} ) .
	\label{eq:SBI_posterior}
\end{equation}

\medskip
Our particular implementation of the SBI pipeline may be summarized via the following steps:
\begin{itemize}
    \item A convolutional neural network is trained to obtain the desired summary, i.e. cluster mass $\tilde{d}$, from the input data, i.e. the 3D phase-space distribution $\myvec{d} \equiv \{ \myvec{x}_{\mathrm{proj}}, \myvec{y}_{\mathrm{proj}}, \myvec{v}_{\mathrm{los}} \}$, using a training set;
    \item A separate test set of simulated clusters is fed to the trained neural network to obtain the corresponding cluster masses;
    \item A Gaussian kernel density estimator is used to compute the joint probability distribution of the summary-parameter pairs, i.e. $\mathcal{P}( \tilde{d} , M )$ (cf. Fig.~\ref{fig:joint_KDE_SBI});
    \item A slice through the above distribution at the network summary prediction $\tilde{d}_{\mathrm{obs}}$, for a given input observation $\myvec{d}_{\mathrm{obs}}$, yields the approximate posterior predictive distribution $\mathcal{P}( M | \myvec{d}_{\mathrm{obs}}, \hat{\myvec{\theta}}, \hat{\myvec{\gamma}} )$.
\end{itemize}

\medskip
The above approach has several key advantages. Although the posterior is conditional on the (trained) network weights and choice of hyperparameters, this SBI framework is guaranteed to provide us with a posterior of the parameter of interest with statistically consistent uncertainties. If the performance and efficacy of the neural network are sub-optimal, as a result of not converging to the global optimal solution, then the uncertainties will only be inflated but not underestimated. Moreover, the density estimator can be precomputed, such that any slice through the likelihood can be computed almost instantaneously to yield the desired approximate posterior for any given observation. In this work, we make use of a Gaussian KDE (cf. Fig.~\ref{fig:joint_KDE_SBI} in Section~\ref{validation_performance}) as our density estimator. The main caveats of this framework are that there is a choice of density estimator with some hyperparameters, such as the bandwidth for the Gaussian KDE, and as for any other SBI approaches, there is some dependence on the total number of simulations used to compute an approximation of the likelihood. Nevertheless, the Gaussian KDE is a fairly robust option as it is not very sensitive to the choice of hyperparameters. In contrast, sophisticated neural density estimators would require further (unsupervised) training and hyperparameter tuning, and would be prone to the shortcoming related to the training of conventional neural networks, i.e. convergence to local minima on the likelihood surface, as outlined above.

\subsection{Convolutional neural networks}
\label{CNN_intro}

Convolutional neural networks (hereafter CNNs) \citep{lecun1995convolutional, lecun1998gradient} are a particular type of artificial neural network, especially suited for problems where spatially correlated information is crucial. In essence, a CNN is designed as follows: A convolutional kernel, commonly referred to as a {\it filter}, of a given size, encoding a set of neurons, is applied to each pixel (or voxel for 3D inputs) of the input image and its vicinity as it scans through the whole region. A given pixel in a specific layer is only a function of the pixels in the preceding layer which are enclosed within the window defined by the kernel, known as the {\it receptive field} of the layer. This yields a {\it feature map} which encodes high values in the pixels which match the pattern encoded in the weights and biases of the corresponding neurons in the convolutional kernel. These weights and biases are the trainable parameters that are optimized during training.

\medskip
To extract the series of distinct features of the input image, a convolutional layer generally employs several filters, resulting in a set of feature maps which are then fed as inputs to the subsequent layer. This convolutional operation is typically followed by a {\it pooling} layer as a subsampling or dimensionality reduction step \citep{goodfellow2016deep}. The application of these two types of layers will reduce the initial input image to a compact representation of features, which can be reshaped as a vector. This feature vector is subsequently passed to the final layer which is a fully connected layer to ultimately generate an output \citep{lecun2015deep}.

\medskip
In terms of the mathematical formalism, the convolutional operation may be described as a specialized linear operation, with the discrete convolution implemented via matrix multiplication. As such, a particular convolutional layer, denoted by $\ell$, can be computed using
\begin{equation}
	x_j^\ell = \mathpzc{F} \left( \sum_{i \in \mathcal{M}_j} x_i^{\ell-1} \times k_{ij}^\ell + b_j^\ell \right) ,
	\label{eq:convolutional_operation}
\end{equation}
where $\mathpzc{F}$ denotes the activation function, $k$ represents the convolutional kernel, $\mathcal{M}_j$ corresponds the receptive field and $b$ is the bias parameter \citep{goodfellow2016deep}. The role of the activation function is to encode some non-linearity in the convolutional layers, so that a stack of such layers can be used as a generic function approximator. We make use of the rectified linear unit ({\ttfamily ReLU}) activation function \citep{nair2010ReLU} in our neural network, as described in Section~\ref{CNN_architecture}, defined as follows:
\begin{align}
	\mathpzc{f}(z_i) = \Bigg\{\begin{matrix}
    0, \: \: \: \: &z_i < 0 \\ z_i , \: \: \: \: &z_i \geq 0. \end{matrix}
	\label{eq:relu_activation}
\end{align}
The \texttt{ReLU} activation and its variants are less computationally expensive than other common activation functions, such as the sigmoid and hyperbolic tangent (\texttt{tanh}) functions, and mitigates the vanishing gradient issue in training deep neural networks. The latter predicament arises when neurons saturate due to an activation function where $\mathpzc{f}(z_i) \approx 0$ or $\mathpzc{f}(z_i) \approx 1$, as in the case of the sigmoid and \texttt{tanh} functions, such that the gradient tends to zero. This is inevitably detrimental to the effectiveness of gradient descent during training, resulting in poor training performance. In our neural network implementation, the \texttt{ReLU} function also ensures positivity of the final output, i.e. the predicted cluster mass.

\medskip
The above CNN design allows the neural network to autonomously extract meaningful spatial features from the input image. By stacking several convolutional layers, the network is capable of building an internal hierarchical representation of features encoding the most relevant information from the input image, such that the network is able to identify increasingly complex patterns with the addition of more layers. Hence, convolutional layers provide a natural approach to take spatial context into consideration. A key aspect of such networks is that a stack of convolutional layers increases the sensitivity of subsequent layers to features on increasingly larger scales. In other words, the size of the receptive field becomes larger as we go deeper in the network. Moreover, convolutional layers retain the local information while performing the convolution on adjacent pixels, thereby allowing both local and global information to propagate through the network \citep{lecun2015deep}.

\medskip
CNNs have recently been developed for a range of cosmological applications involving the distribution of cosmic structures on various scales. Deep CNNs, based on the U-Net model \citep{ronneberger2015UNet}, have been designed to predict the non-linear cosmic structure formation from linear perturbation theory \citep{he2019learning} and to include physical effects induced by the presence of massive neutrinos in standard dark matter simulations \citep{giusarma2019learning}. \citet{zhang2019darkmatter} devised a two-phase CNN architecture to map 3D dark matter fields to their corresponding galaxy distributions in hydrodynamic simulations. 3D deep CNNs have also been used for the generation of mock halo catalogues \citep{berger2018volumetric, bernardini2019predicting} or for the classification of the distinct features of the cosmic web from $N$-body simulations \citep{aragon2019classifying}. Physically motivated CNNs, based on the Inception architecture \citep{szegedy2017residualinception}, have been constructed to map 3D dark matter fields to their halo count distributions \citep{DKR2019painting} and to augment low-resolution $N$-body simulations with high-resolution structures \citep{DKR2020super}.

\subsection{Neural network architecture}
\label{CNN_architecture}

\begin{figure*}
	\centering
		{\includegraphics[width=\hsize,clip=true]{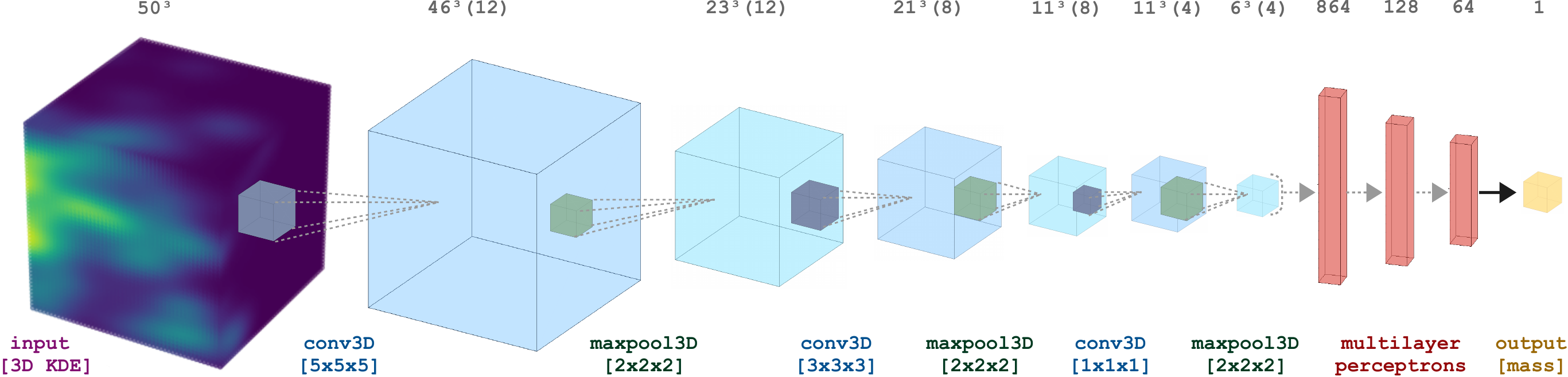}}
	\caption{Schematic representation of our CNN$_{\mathrm{3D}}$ architecture to predict cluster masses from their 3D dynamical phase-space distributions. The dimensions of the input 3D slice and those of the subsequent slices, resulting from the convolutional and maxpooling operations, are indicated in the top row, with the number of feature maps per layer given in parentheses. The respective kernel sizes of the latter operations are given in the bottom row, with single strides employed and without use of padding. The CNN extracts the informative spatial features from the 3D phase-space distribution and gradually compresses the high-dimensional space to a single scalar which corresponds to the dynamical cluster mass.}
	\label{fig:CNN_schematic}
\end{figure*}

The underlying objective of our SBI framework is to infer the posterior of the dynamical mass of a galaxy cluster, given its 3D phase-space distribution characterized by the projected sky positions and the line-of-sight velocities, i.e. $\mathcal{P}(M | \{ \myvec{x}_{\mathrm{proj}}, \myvec{y}_{\mathrm{proj}}, \myvec{v}_{\mathrm{los}} \})$. The neural network takes as input a 3D slice $\Dtilde$, which is a 3D array of the Gaussian KDE applied to the phase-space distribution, as described in Section~\ref{dynamical_3D_phase_space_distribution}, with an example illustrated in Fig.~\ref{fig:phase_space_distribution}. The training data set, therefore, consists of pairs of $\{ M, \Dtilde \}$.

\medskip
A schematic of our 3D CNN (hereafter CNN$_{\mathrm{3D}}$) architecture is depicted in Fig.~\ref{fig:CNN_schematic}. The network extracts spatial features from the input 3D phase-space distribution by performing convolutions with a kernel of size $5\times5\times5$. We employ several such kernels in one layer to probe different aspects of the input 3D slice, yielding a set of feature maps which are subsequently fed to a maxpooling layer for the purpose of dimensionality reduction. We use a $2\times2\times2$ maxpooling kernel to reduce the slice size by a factor of two. We adopt single strides and no padding for both operations. By repeatedly alternating between these two types of layers, we can reduce the initial 3D distribution to a compact representation of features. At this point, the resulting 3D slice may be flattened to a vector, with this vectorized set of features fed to the final layers which consist of fully connected layers of neurons, i.e. multilayer perceptrons. Finally, the output layer yields the dynamical cluster mass, as desired. We encode \texttt{ReLU} \citep{nair2010ReLU} activation functions in the convolutional layers, and linear activations in the final fully connected layers. We highlight the relatively low complexity of the network architecture with $\sim 10^5$ trainable weights.

\subsection{Training methodology}
\label{training_methodology}

We train our CNN$_{\mathrm{3D}}$ model as a regression over the logarithmic cluster mass by minimizing a mean squared error loss function with respect to the network weights. The model and training routine are implemented using the \textsc{Keras} library \citep{chollet2015keras} via a \textsc{TensorFlow} backend \citep{abadi2016tensorflow}. We make use of the {\it Adam} \citep{kingma2014adam} optimizer, with a learning rate of $\eta=10^{-4}$ and first and second moment exponential decay rates of $\beta_1=0.9$ and $\beta_2=0.999$, respectively. The batch size is set to 100. We train the neural network for around 50 epochs, requiring around 10 minutes on an NVIDIA V100 Tensor Core GPU. In order to prevent any overfitting, we adopt the standard regularization technique of early stopping in our training routine. For this purpose, $25\%$ of the original training data set is kept as a separate validation set, with both the training and validation losses monitored during training. We opt for an early stopping criterion of 5 epochs, such that training is halted when the validation loss no longer shows any improvement for 5 consecutive epochs, and the optimized weights of the previously saved best fit model are restored.

\section{Validation and performance}
\label{validation_performance}

\begin{figure}
	\centering
    \includegraphics[width=\hsize, clip=true]{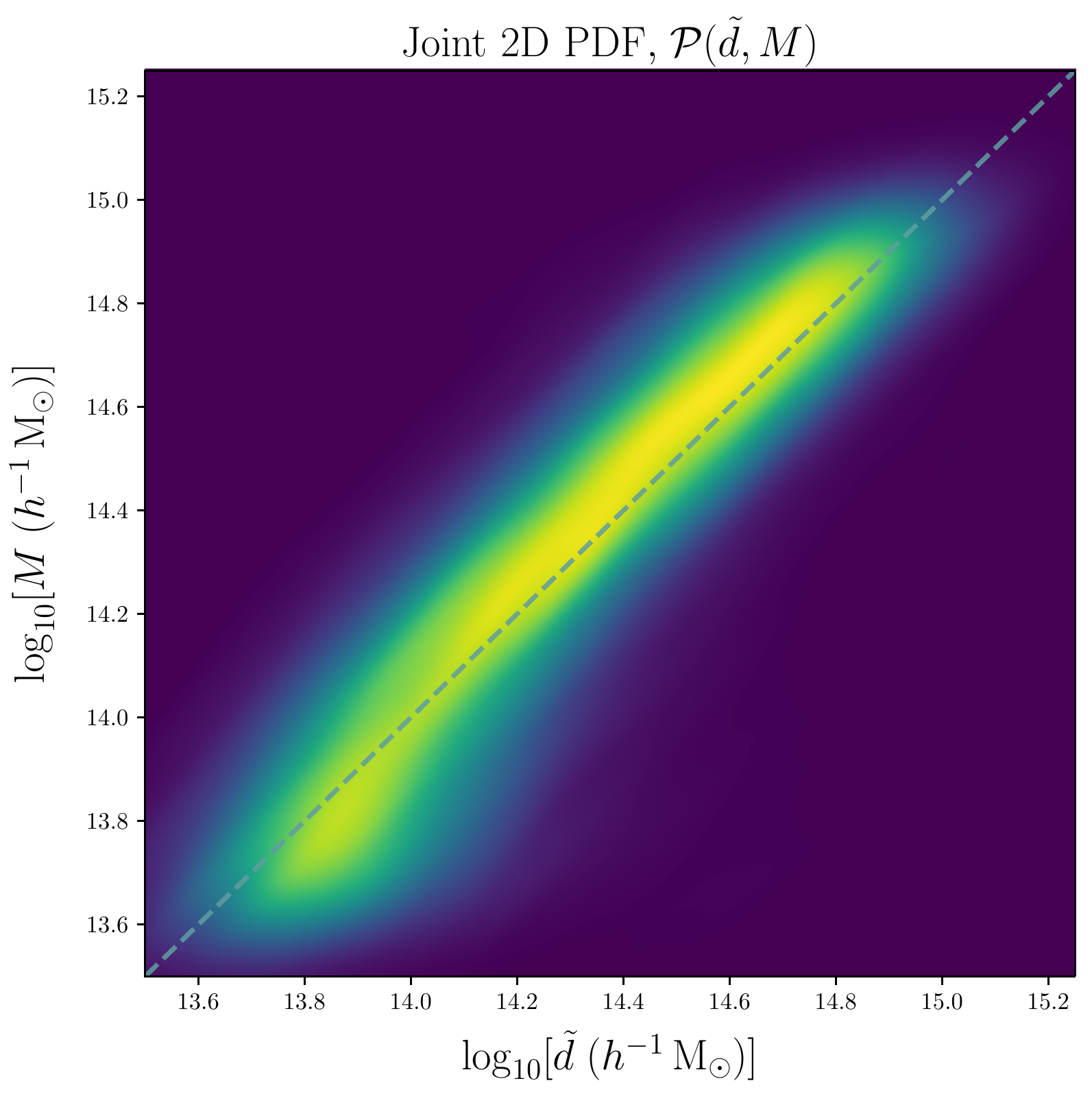}
	\caption{Joint 2D PDF of network predicted summaries $\tilde{d}$ and parameters $M$, i.e. $\mathcal{P}(\tilde{d}, M)$, obtained using a bivariate Gaussian KDE with a bandwidth scaling factor of $0.20$. Recall that $\tilde{d}$, by design, corresponds to the point dynamical mass estimates from our CNN$_{\mathrm{3D}}$ model. A test set consisting of twenty thousand clusters is used to compute this 2D PDF. This is representative of the prediction scatter with respect to the ground truth masses and is employed in our simulation-based inference framework to compute and assign uncertainties associated to point masses predicted by our CNN$_{\mathrm{3D}}$. We obtain the approximate posterior, $\mathcal{P}(M | \{ \myvec{x}_{\mathrm{proj}}, \myvec{y}_{\mathrm{proj}}, \myvec{v}_{\mathrm{los}}  \}, \myvec{\theta}, \myvec{\alpha})$, given a set of network weights $\myvec{\theta}$ and hyperparameters $\myvec{\alpha}$, for a particular cluster by making a vertical slice at the neural network predicted value of $\tilde{d}$.}
	\label{fig:joint_KDE_SBI}
\end{figure}

\begin{figure}
	\centering
    \includegraphics[width=\hsize, clip=true]{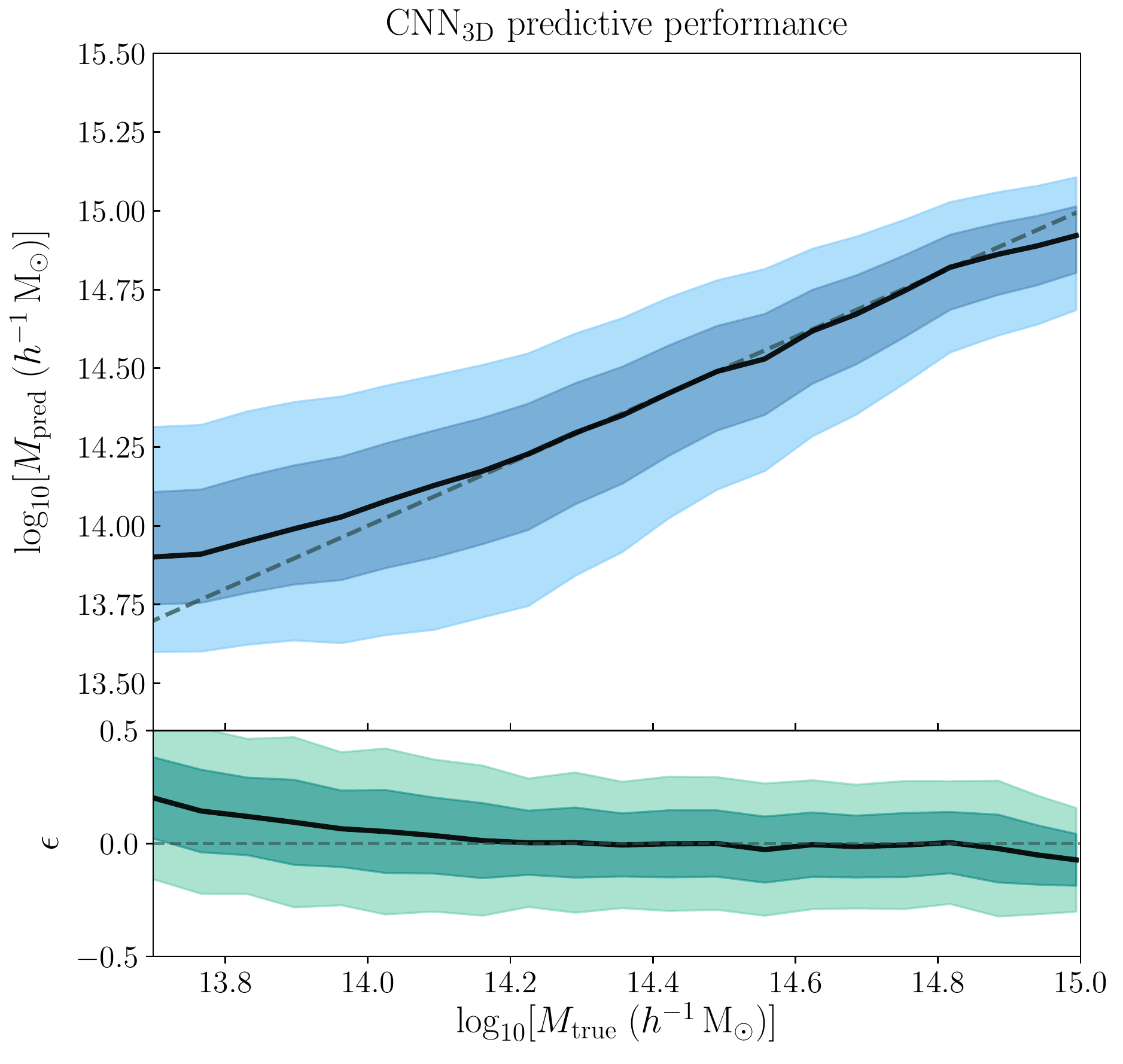}
	\caption{Predictive performance of our CNN$_{\mathrm{3D}}$ model. {\it Top panel:} CNN$_{\mathrm{3D}}$ predictions against ground truth, depicting the mean prediction (solid line) and the predicted confidence intervals (shaded $1\sigma$ and $2\sigma$ regions) of the posterior probability density as a function of logarithmic bins of $M_{\rm true}$ for $\sim 5000$ galaxy clusters from the evaluation set. The simulation-based inference approach, as expected, yields larger uncertainties for low-mass clusters.
	{\it Bottom panel:} Distribution of residual scatter as a function of the logarithmic true cluster mass. The solid line corresponds to the mean logarithmic residual scatter, $\epsilon \equiv \log_{10} (M_{\mathrm{true}} / M_{\mathrm{pred}})$, if we consider only the maximum likelihood predictions (i.e. point mass estimates) from our CNN$_{\mathrm{3D}}$, in logarithmic bins of $M_{\rm true}$. The shaded bands depict the log-normal scatter ($1\sigma$ and $2\sigma$ regions) about the mean residuals. The CNN$_{\mathrm{3D}}$ tends to overestimate masses of poor clusters below $\log[M_{\rm true} (h^{-1} {\rm M}_{\odot})] \approx 14.0$~dex. The correspondingly larger uncertainties for clusters in this mass regime demonstrate the reliability of the simulation-based inference framework to provide uncertainties that are not underestimated.}
	\label{fig:combined_cluster_mass_predictions}
\end{figure}

\subsection{Uncertainty estimation}
\label{uncertainty_estimation}

We now assess the performance of our optimized CNN$_{\mathrm{3D}}$ model on the evaluation set. As part of the SBI procedure, as described in Section~\ref{SBI_sub}, we first compute the joint 2D probability density function (PDF), $\mathcal{P}(\tilde{d}, M)$, of the summary statistics $\tilde{d}$ extracted by the neural network and the parameters $M$ obtained using the test set containing around twenty thousand clusters (cf. Section~\ref{mock_catalogues}). Recall that the neural summary statistics in this case are, by design, taken to be point predictions of masses by the CNN$_{\mathrm{3D}}$, while the parameters correspond to the ground truth masses. We make use of a bivariate Gaussian KDE, with a bandwidth scaling of $\kappa = 0.20$, to obtain the 2D PDF depicted in Fig.~\ref{fig:joint_KDE_SBI}. This involves the application of equations~\eqref{eq:KDE_definition} and \eqref{eq:kernel_definition}, where now $\myvec{x} = (\tilde{d}, M)$ corresponds to a given evaluation point in this 2D parameter space and $\myvec{X}_i$ describes the collection of pairs of $\{ \tilde{d}, M \}$ data points, with $\mathbf{H}$ being a $2\times2$ bandwidth matrix. To infer the posterior PDFs of the dynamical masses of the clusters in the evaluation set, we first obtain the network point predictions using the trained CNN$_{\mathrm{3D}}$ model. We subsequently vertically slice the joint PDF from Fig.~\ref{fig:joint_KDE_SBI} at the point estimates to infer the approximate posterior PDFs for the mass of each cluster. From the posteriors, we quantify the $1\sigma$ uncertainties by integrating the $68\%$ probability volume, such that the upper and lower $1\sigma$ uncertainty limits may be asymmetrical.

\subsection{Performance evaluation}
\label{neural_network_evaluation}

Using the inferred posterior mass PDFs for the clusters in the evaluation set, we evaluate the performance of our CNN$_{\mathrm{3D}}$ on the realistic mock catalogue by plotting our model predictions against the ground truth masses of the $\sim 5000$ clusters from the evaluation set in the top panel of Fig.~\ref{fig:combined_cluster_mass_predictions}. We bin the model predictions in logarithmic mass intervals with the mean prediction and confidence intervals ($1\sigma$ and $2\sigma$ regions) of the posterior probability density depicted via the solid line and shaded regions, respectively. The top panel shows the efficacy of our CNN$_{\mathrm{3D}}$ model to recover the ground truth masses of the clusters from the evaluation set within the $1\sigma$ uncertainty limit. The bottom panel displays the distribution of residuals, $\epsilon \equiv \log_{10} (M_{\mathrm{true}} / M_{\mathrm{pred}})$, in the CNN$_{\mathrm{3D}}$ point predictions relative to the ground truth, as a function of the logarithmic cluster mass, with the solid line indicating the mean residual scatter and the shaded bands corresponding to the $1\sigma$ and $2\sigma$ regions. The CNN$_{\mathrm{3D}}$ predictions have a mean residual and log-normal scatter of $\langle \epsilon \rangle = 0.04$~dex and $\sigma_{\epsilon} = 0.16$~dex.

\medskip
From the bottom panel of Fig.~\ref{fig:combined_cluster_mass_predictions}, we observe the tendency of the CNN$_{\mathrm{3D}}$ to overestimate the masses for clusters with masses below $\log[M_{\rm true} (h^{-1} {\rm M}_{\odot})] \approx 14.0$~dex. This may primarily be attributed to the realistic effects included in our mock catalogue as detailed in Section~\ref{interloper_visualization} below. Nevertheless, this relatively high residual scatter due to the overprediction of cluster mass is properly accounted for in the network predicted uncertainties, with the lower $1\sigma$ and $2\sigma$ limits being larger than the upper limits. This demonstrates the capacity of the SBI framework to yield reliable uncertainties that are not underestimated. Conversely, the network slightly underestimates the masses for the most massive clusters above $\log[M_{\rm true} (h^{-1} {\rm M}_{\odot})] \approx 14.9$~dex. There is a two-fold plausible explanation for this effect. First, the selection cuts (cf. Section~\ref{mock_catalogues}) to produce the 3D phase-space diagrams may not be sufficiently large to capture all the galaxy members of the massive clusters, resulting in incomplete cluster samples. The second explanation is related to possible mean-reversion edge effects, as also reported by \citet{ntampaka2016dynamical, ho2019robust, ho2020approximate}, whereby the model predictions of cluster masses at the edge of the mass range considered here are biased towards the average. In general, this systematic bias is related to a neural network's tendency to be more adept at interpolation than extrapolation. To mitigate such biases, we would require more training clusters beyond the edges of the mass regime of the training set, i.e. $\log[M_{\rm true} (h^{-1} {\rm M}_{\odot})] < 13.7$~dex and $\log[M_{\rm true} (h^{-1} {\rm M}_{\odot})] > 15.0$~dex. Note that this mean-reversion effect may also be partially responsible for the overprediction of cluster masses in the low-mass regime.

\begin{figure}
	\centering
    \includegraphics[width=\hsize, clip=true]{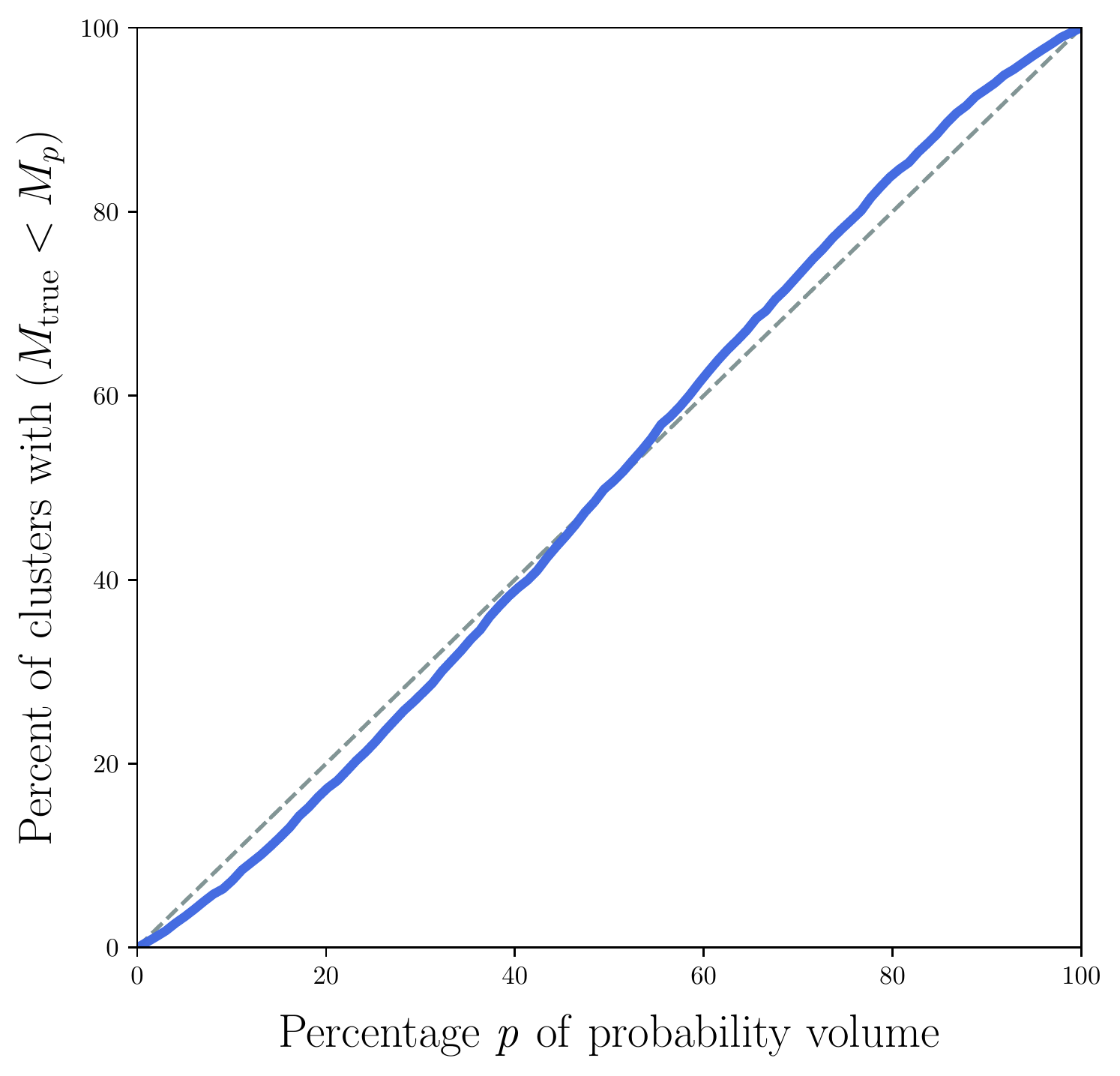}
	\caption{Validation of posterior recovery within the SBI framework via a posterior calibration (or quantile-quantile) plot. The dashed line indicates perfect or ideal calibration, with the solid blue line depicting the calibration of the inferred posterior PDFs. The latter illustrates the overall statistical consistency of the inferred posteriors, displaying near ideal calibration, thereby avoiding both under and overconfidence.}
	\label{fig:posterior_validation}
\end{figure}

\medskip
For the validation of posterior recovery within the SBI framework, we perform a similar statistical test to recent ML studies \citep[e.g.][]{perrault2017uncertainties, ho2020approximate, wagnercarena2020hierarchical}. This test is based on model calibration, whereby a posterior containing $x\%$ of the probability volume should contain the ground truth within this specific volume $x\%$ of the time. As such, for our given problem with one-dimensional mass posteriors, this test boils down to the computation of coverage probabilities, defined as the fraction of test samples where the ground truth lies within a particular confidence interval. Note that this posterior validation test does not assume Gaussian nature of PDFs and holds for any arbitrary PDFs. A clear and intuitive description of this test on a toy model is provided in the appendix of \citet{wagnercarena2020hierarchical}. The posterior calibration plot for the $\sim 5000$ clusters in the evaluation set, averaged over all clusters, is depicted in Fig.~\ref{fig:posterior_validation}. The diagonal dashed line implies ideal calibration, as a result of a perfect match between the number of test samples and the percentage of probability volume. The calibration plot of the inferred posterior PDFs is depicted via the solid blue line, illustrating the near ideal calibration and overall statistical consistency of the inferred posteriors. The absence of any significant deviations from the ideal calibration reference line implies that the SBI framework does not exhibit any particular strong under or overconfidence in assigning uncertainties to the CNN$_{\mathrm{3D}}$ model predictions.

\subsection{Visualization of interloper contamination}
\label{interloper_visualization}

Our mock catalogue contains a realistic level of contamination by interloper galaxies, as expected from the actual SDSS observations. In this section, we explicitly highlight how the presence of these spurious galaxies renders the mass estimation extremely challenging.

\medskip
The interloper contamination principally induces a bias (overestimation) in the neural network predictions which is more significant for the low-mass clusters, substantiating the relatively larger residual scatter for clusters with masses smaller than $\sim 14.0$~dex, as depicted in the bottom panel of Fig.~\ref{fig:combined_cluster_mass_predictions}. This outcome is caused by several low-mass clusters, for which the CNN$_{\mathrm{3D}}$ systematically and significantly overpredicts the mass. To illustrate that interlopers constitute the underlying cause of these inaccurate mass estimates, we compute the contamination per cluster as the mass ratio of interloper clusters to the original cluster. A cluster is considered to be an interloper cluster when it is more massive than the original cluster, is located within a distance of $R_{\mathrm{proj}} = (x_{\mathrm{proj}}^2 \, + \, y_{\mathrm{proj}}^2)^{1/2} = 4 \; h^{-1}$ Mpc and has $\Delta v_{\rm{los}} < 2200 \; \rm{km} \; \rm{s}^{-1}$. An additional factor that exacerbates this problem and renders the task of the CNN$_{\mathrm{3D}}$ more convoluted is the distance between the interloper and original clusters in 3D phase space. The closer the two clusters are together, the more difficult it is to tell them apart. The relative phase-space distance between the two clusters, denoted by $c_1$ and $c_2$, respectively, is computed as
\begin{align} \label{eq:3Dphasespace}
&d(c_1, c_2) = \sqrt{ \left( \frac{\Delta x_{\mathrm{proj}}}{R_{200c, \textrm{av}}} \right)^2 + \left( \frac{\Delta y_{\mathrm{proj}}}{R_{200c, \textrm{av}}} \right)^2  + \left( \frac{\Delta v_{\mathrm{los}}}{v_{200c, \textrm{av}}} \right)^2 }, \\
&\textrm{with} \ R_{200c} = \left( \frac{3 M_{200c}}{4 \pi 200 \rho_c} \right)^{1/3} \ \textrm{and} \ \, v_{200c} = \sqrt{ \frac{GM_{200c}}{R_{200c}} }, \nonumber
\end{align}
where $R_{200c, \textrm{av}} = \tfrac{1}{2}(R^{c_1}_{200c} + R^{c_2}_{200c})$ and $\Delta x_{\mathrm{proj}} = x^{c_1}_{\mathrm{proj}} - x^{c_2}_{\mathrm{proj}}$, with $\Delta y_{\mathrm{proj}}$, $\Delta v_{\mathrm{los}}$ and $v_{200c, \textrm{av}}$ analogously defined.

\begin{figure}
	\centering
    \includegraphics[width=\hsize, clip=true]{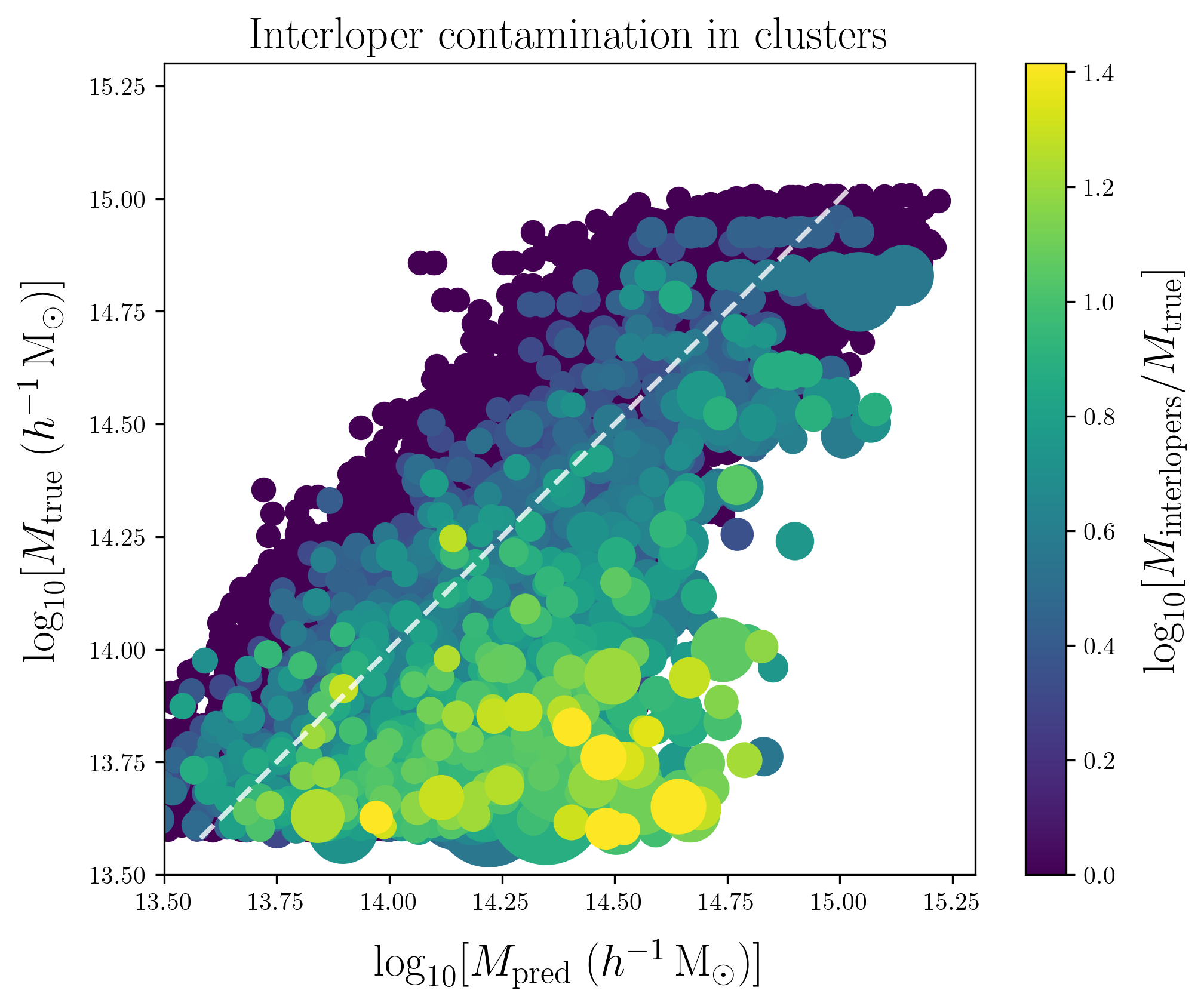}
	\caption{Effect of clustered interlopers on the CNN$_{\mathrm{3D}}$ mass predictions. Each individual mass measurement is coloured according to the relative mass of interloper clusters contaminating the phase-space diagram of the main cluster. In addition, the size of the markers indicates the inverse distance between the interloper and original clusters in the projected phase space, as defined by equation~\eqref{eq:3Dphasespace}. Interlopers residing in relatively massive clusters which overlap closely with the original cluster in the projected phase space can hardly be distinguished from the original cluster members, giving rise to a substantial mass overestimation.}
	\label{fig:contamination}
\end{figure}

\medskip
Fig.~\ref{fig:contamination} provides a stark illustration of the overwhelming interloper contamination inherent to the individual clusters from the test set, with the colour bar corresponding to the degree of interloper contamination and the marker size corresponding to the inverse of the distance. As can be seen, the clusters with the largest overestimation of their dynamical masses are also the ones whose phase-space diagrams are highly contaminated by interlopers in the form of independent clusters. For some of these clusters, the interloper cluster is around 20 times more massive than the original cluster, which renders the mass estimation extremely challenging. Observational data, such as the SDSS catalogue used in this work, would be similarly plagued by interloper contamination. While the performance of our CNN$_{\mathrm{3D}}$ model with a mean residual and log-normal scatter of $\langle \epsilon \rangle = 0.04$~dex and $\sigma_{\epsilon} = 0.16$~dex is not as impressive as the recent ML techniques at first glance, this is purely due to the more realistic mock catalogue employed here.

\subsection{Information gain with higher dimensionality}
\label{info_gain_dimensionality}

\begin{figure*}
	\centering
    \includegraphics[width=\hsize, clip=true]{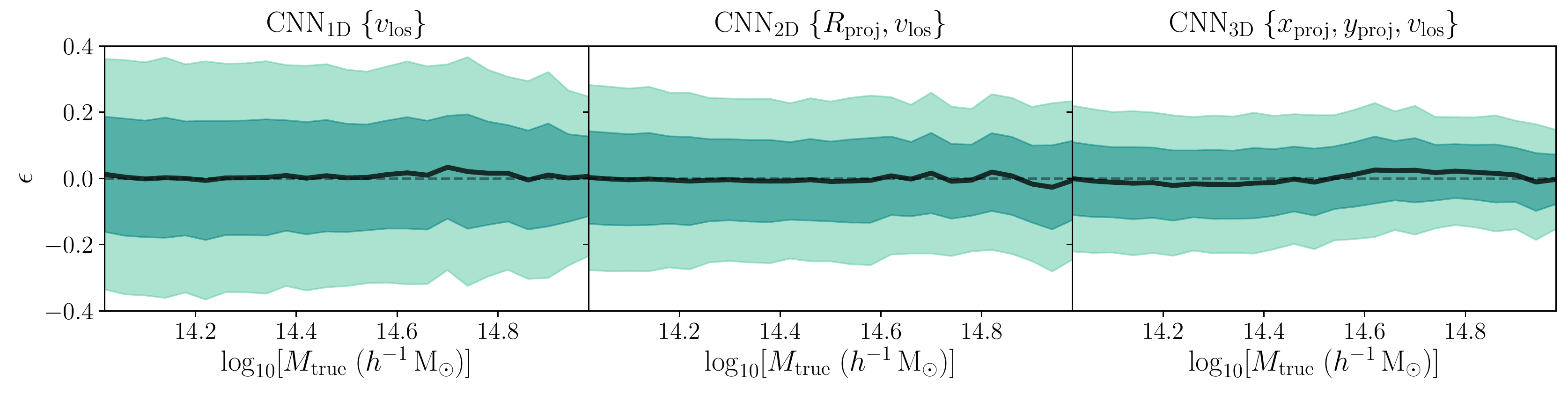}
	\caption{Comparison of the performance of three CNN models with distinct input dimensionality. Performance is quantified in terms of the residual scatter, $\epsilon \equiv \log_{10} (M_{\mathrm{true}} / M_{\mathrm{pred}})$, in the CNN point predictions relative to the ground truth. The mean residual scatter is depicted via solid dark lines, with the shaded bands corresponding to the log-normal scatter ($1\sigma$ and $2\sigma$ regions). The CNNs are trained with progressively larger dimensionality of the phase-space distribution of the same mock cluster catalogue from \citet{ho2019robust}. In all cases, the adopted maximum projected distance from the cluster centre is $1.6\,h^{-1}\rm{Mpc}$. The results illustrate the gain in constraining power when the information content of the full 3D phase-space distribution of galaxies is exploited instead of relying merely on the velocity dispersion as in the 1D case displayed in the left panel. The CNN$_{\mathrm{1D}}$ and CNN$_{\mathrm{2D}}$ results are reproduced from \citet{ho2019robust}.}
	\label{fig:CNN_comparison_plot}
\end{figure*}

\begin{figure}
	\centering
    \includegraphics[width=\hsize, clip=true]{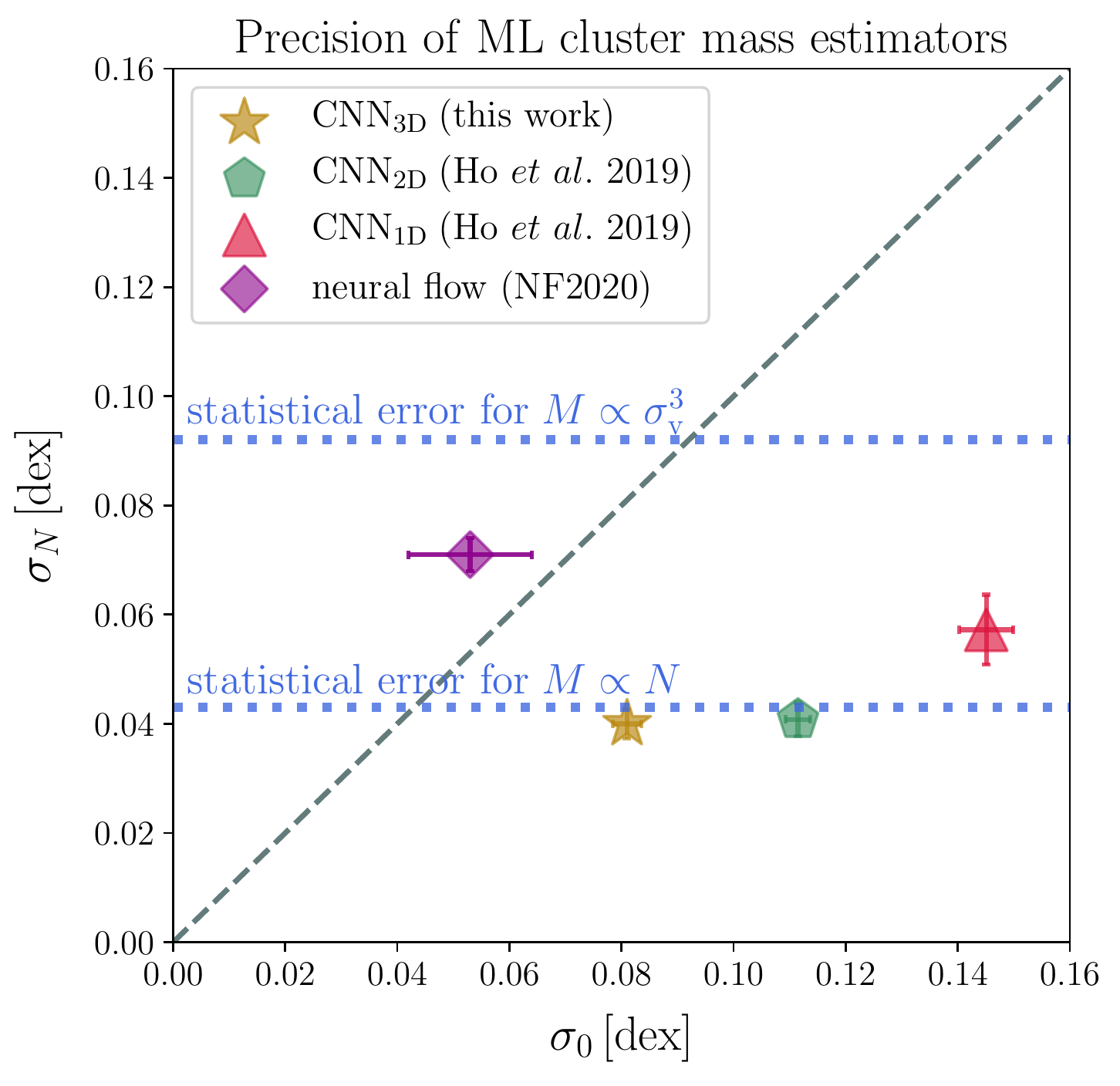}
	\caption{Comparison of the precision of three recently proposed ML cluster mass estimators, along with the CNN$_{\mathrm{3D}}$ model from this work, all based on mock observations from \citet{ho2019robust} with the maximum projected distance from the cluster centre of $1.6\,h^{-1}\rm{Mpc}$. We quantify the precision in terms of richness-dependent error $\sigma_{N}$ and richness-independent systematic error $\sigma_{0}$ (cf. equation~\eqref{eq:errors}). In accordance with Fig.~\ref{fig:CNN_comparison_plot}, this shows the progressive improvement in the precision of CNN models with increasing input dimensionality. The horizontal dotted lines indicate the two characteristic levels of Poisson-like scatter for velocity dispersion-based and richness-based methods. Our CNN$_{\mathrm{3D}}$ model is also less sensitive to the cluster richness than the neural flow mass estimator (NF2020).}
	\label{fig:CNN_precision_plot}
\end{figure}

In an attempt to illustrate the gain in information by exploiting the full 3D phase-space distribution, we also train our CNN$_{\mathrm{3D}}$ on the mock catalogue from \citet{ho2019robust} and compare the performance of our network to their 1D and 2D counterparts in terms of the logarithmic residual scatter in Fig.~\ref{fig:CNN_comparison_plot}. CNN$_{\mathrm{1D}}$ infers cluster masses solely from the univariate distribution of line-of-sight velocities, i.e. $\{ v_{\mathrm{los}} \}$, while CNN$_{\mathrm{2D}}$ additionally takes as input the sky-projected radial positions given by $R_{\mathrm{proj}} = (x_{\mathrm{proj}}^2 + y_{\mathrm{proj}}^2)^{1/2}$, such that it relies on the joint distribution of $\{ R_{\mathrm{proj}}, v_{\mathrm{los}} \}$. For the sake of comparison, we use similar phase-space cuts to \citet{ho2019robust} for our CNN$_{\mathrm{3D}}$ model, i.e.  $v_{\mathrm{los}} \in [-2200, 2200] \, \rm{km} \, \rm{s}^{-1}$, $x_{\mathrm{proj}} \in [-1.6, 1.6] \, h^{-1} \rm{Mpc}$ and $y_{\mathrm{proj}} \in [-1.6, 1.6] \, h^{-1} \rm{Mpc}$. Note that only the point (maximum {\it a posteriori}) estimates from our approach are used to produce the comparison plot displayed in Fig.~\ref{fig:CNN_comparison_plot}. As expected, we find that the precision of the mass estimator, as indicated by the log-normal residual scatter (shaded $1\sigma$ and $2\sigma$ regions), improves progressively with further information, thereby justifying the development and application of our CNN$_{\mathrm{3D}}$ model in this work.

\medskip
As in our previous work (NF2020), we quantify the precision of cluster mass estimation in terms of the total scatter about the best-fit power-law relation between the ground truth and predicted cluster masses. Adopting the approach employed in \citet{wojtak2018clusterIV}, we express the total scatter $\sigma$ into a richness-dependent component given by $\sigma_{N}$ and a richness-independent part denoted by $\sigma_{0}$, as follows:
\begin{equation}
\sigma^{2} = \sigma_{N}^{2}(N_{\rm mem}/100)^{-1} + \sigma_{0}^{2},
\label{eq:errors}
\end{equation}
where $N_{\rm mem}$ indicates the number galaxies within $R_{200c}$ of the cluster's host dark matter halo. We determine the values of $\sigma_{N}$ and $\sigma_{0}$ by fitting the above equation to the logarithmic residuals in the cluster mass predictions for the test set. The best fit model recovers the measured scatter with a fully satisfactory precision of 5 per cent. We carry out this procedure for the three CNN models and also include the results of the neural flow mass estimator (NF2020). Note that the same mock cluster catalogue from \citet{ho2019robust} was used for all the methods. The recent ML techniques, illustrated in Fig.~\ref{fig:CNN_precision_plot}, all outperform the traditional cluster mass estimators (cf. Fig.~6 in NF2020) extensively tested in the Galaxy Cluster Mass Comparison Project \citep{old2015clusterII}, which are not shown for the sake of clarity. For our CNN$_{\mathrm{3D}}$ model, we find $\sigma_{N}=0.04$~dex and $\sigma_{0}=0.08$~dex, with the richness-dependent error smaller by a factor of two relative to $3/(\sqrt{2N_{\rm mem}}\ln10)=0.09$~dex expected for the mass estimation based solely on the scaling relation with the velocity dispersion. As expected from Fig.~\ref{fig:CNN_comparison_plot}, Fig.~\ref{fig:CNN_precision_plot} shows a progressive improvement in precision going from CNN$_{\mathrm{1D}}$ to CNN$_{\mathrm{3D}}$ due to the gain in constraining power when exploiting the full information from the 3D phase-space distribution of galaxies rather than relying only on the velocity dispersion. In general, the CNN mass estimators are less sensitive to cluster richness than the neural flow model. Figs.~\ref{fig:CNN_comparison_plot} and \ref{fig:CNN_precision_plot}, therefore, present an adequate depiction of the network performance and a fair comparison with recent ML methods, demonstrating the precision of our CNN$_{\mathrm{3D}}$ model.

\section{Application to SDSS catalogue}
\label{applications_sdss}

\begin{figure}
	\centering
    \subfloat{\includegraphics[width=0.825\hsize]{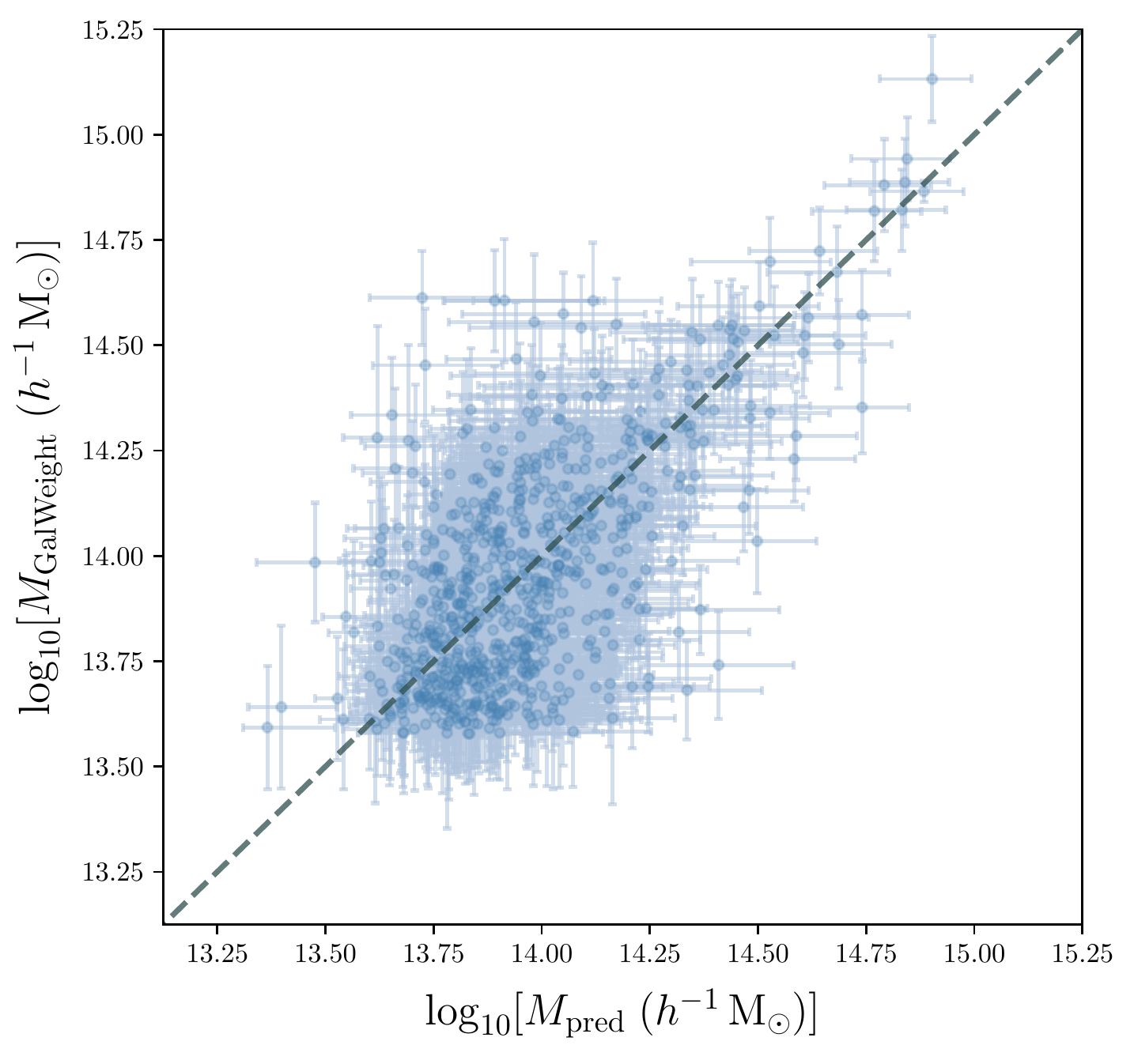}}
    \quad
    \subfloat{\includegraphics[width=0.8\hsize, clip=true]{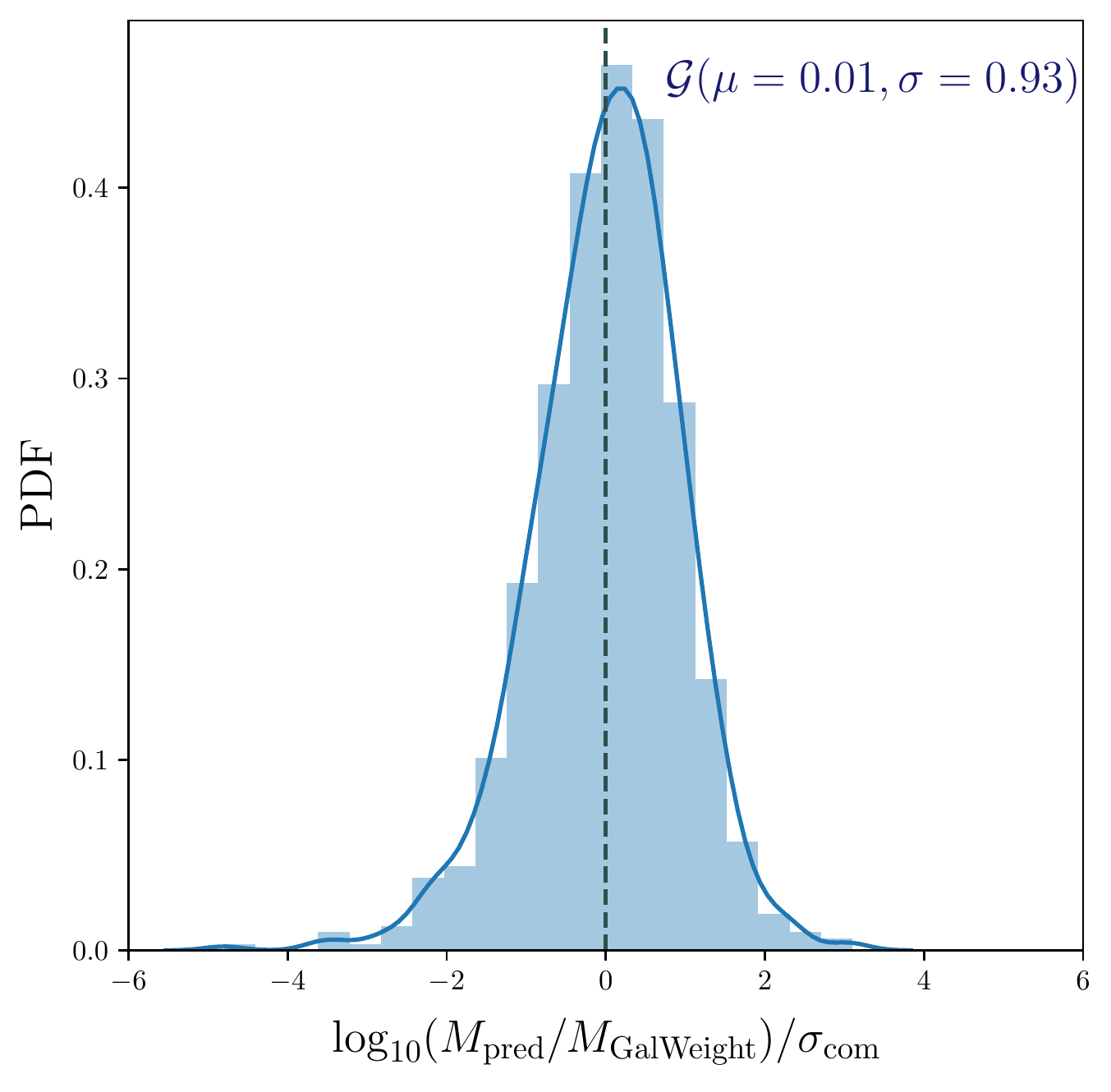}}
    \quad
	\caption{Comparison of our SDSS cluster mass predictions with the recent estimates from the {\it GalWeight} galaxy cluster catalogue \citep{Abdullah2020}, illustrated via a qualitative visual depiction ({\it top panel}) and a 1D PDF ({\it bottom panel}) of the difference between the predictions, normalized by the corresponding uncertainties in our predictions. The resulting distribution is approximately characterized by a normalized Gaussian distribution, quantitatively indicating the overall consistency between our cluster mass predictions and those from \citet{Abdullah2020}.}
	\label{fig:mass_contrast_PDF}
\end{figure}

We now apply the trained neural network to redshift data from the SDSS catalogue to infer the dynamical masses of the galaxy clusters and use the bivariate KDE (cf. Fig.~\ref{fig:joint_KDE_SBI}) to derive their corresponding uncertainties. We subsequently perform a detailed comparison of the inferred dynamical masses to recent measurements from the literature.

\medskip
We use the publicly available \textit{GalWeight} catalogue containing galaxy clusters found in the main galaxy sample of the SDSS with the \textsc{galweight} algorithm \citep{Abdullah2020}.\footnote{\url{https://mohamed-elhashash-94.webself.net/galwcat/}} We select galaxy clusters at comoving distances shorter than the upper limit assumed in the mock data, i.e. $250\,h^{-1}\,\rm{Mpc}$, and we discard clusters whose observational cones given by the maximum physical distance of $x_{\rm proj}$ and $y_{\rm proj}$ are not included in the footprint of the SDSS main galaxy sample. The resulting sample consists of 801 galaxy clusters. For each cluster, we find velocities and positions of all galaxies from the main spectroscopic SDSS sample in its field. Angular positions were converted into physical distances assuming the deceleration parameter $q_{0}=-0.55$ consistent with the Planck cosmology \citep{13planck2015}, although the impact of cosmological model in the adopted redshift range $(z\leq0.085)$ is negligible. We use the same cuts in the projected phase space coordinates as for the mock observations. We also adopt cluster centres and redshifts from the \textit{GalWeight} catalogue which set them at the peak of a smoothed galaxy density in the projected phase space. The cluster catalogue also provides the measurements of dynamical cluster masses based on the virial theorem with the surface term computed for NFW density profile extrapolated beyond the virial sphere. The mass estimations account for cluster membership by a special scheme of assigning weights to all galaxies observed in the phase-space diagram. The scheme was devised using mock data generated from cosmological simulations \citep{abdullah2018galweight}.

\medskip
The galaxy clusters from the catalogue are subjected to an initial preprocessing step similar to the preparation of the training set. We compute the 3D Gaussian KDE of their respective phase-space distributions, as outlined in Section~\ref{kde}, with the resulting 3D slices subsequently provided as inputs to our CNN$_{\mathrm{3D}}$ model. The point estimates are then fed to the SBI pipeline to obtain their respective uncertainties, resulting in the inferred dynamical masses for the SDSS clusters. To compare our predictions with the recent results from \citet{Abdullah2020}, we compute the 1D PDF of the difference between the two sets of predictions, normalized by the combined uncertainties, i.e. $\log_{10} (M_{\mathrm{pred}} / M_{\mathrm{GalWeight}}) / \sigma_{\mathrm{com}}$, where $\sigma_{\mathrm{com}} \equiv (\sigma_{\mathrm{pred}}^2 + \sigma_{\mathrm{GalWeight}}^2)^{1/2}$ and $M_{\mathrm{GalWeight}}$ is the cluster mass estimate from the \textit{GalWeight} galaxy cluster catalogue \citep{Abdullah2020} with associated uncertainty $\sigma_{\mathrm{GalWeight}}$. We compute the 1D PDF by binning this mass contrast, with the resulting distribution illustrated in Fig.~\ref{fig:mass_contrast_PDF}. The latter distribution has a mean and standard deviation of $\mu = -0.02$ and $\sigma = 1.05$, respectively, which approximately corresponds to a normalized Gaussian distribution. This highlights the overall consistency of our mass predictions with those from \citet{Abdullah2020}, with the absence of kurtosis implying a negligible bias or error underestimation/overestimation with respect to the former literature estimates, which would otherwise render the 1D PDF leptokurtic or platykurtic.

\section{Conclusions and outlook}
\label{conclusions}

We have presented a simulation-based inference framework, based on 3D convolutional feature extractors, to infer the galaxy cluster masses from their 3D dynamical phase-space distributions, which consist of the projected positions in the sky and the galaxy line-of-sight velocities, i.e. $\{ x_{\mathrm{proj}}, y_{\mathrm{proj}}, v_{\mathrm{los}} \}$. The simulation-based inference framework allows us to quantify the uncertainties on the inferred masses in a straightforward and robust way. By optimally exploiting the information content of the full projected phase-space distribution, the network yields dynamical cluster mass estimates with precision comparable to the best existing traditional methods. As such, this fast and robust tool is a novel and complementary addition to the state-of-the-art machine learning techniques in the cluster mass estimation toolbox.

\medskip
We train our CNN$_{\mathrm{3D}}$ model using a realistic mock cluster catalogue emulating the properties of the actual SDSS catalogue. Once optimized on the training set, we use our CNN$_{\mathrm{3D}}$ model within a simulation-based inference framework to infer the dynamical masses of a set of SDSS clusters and their associated uncertainties for the first time using a machine learning-based cluster mass estimator, and we obtain results consistent with mass estimates from the literature. The primary advantage of simulation-based inference, as employed in this work, is that it yields accurate and statistically consistent uncertainties. If the neural network used to perform the feature extraction to derive summary statistics is sub-optimal, the uncertainties will only be inflated, thereby obviating overconfident posteriors. Moreover, we clearly illustrate the difficulties related to the presence of interlopers close to the cluster centre when dealing with actual observations. In practice, there exists no effective solution to this predicament inherent to the mass estimation problem. As a consequence, our simulation-based inference framework yields correspondingly larger uncertainties for such problematic clusters. This conservative approach ensures that the uncertainties of highly contaminated clusters are not underestimated.

\medskip
The design of our network architecture, based on the use of 3D convolutional kernels, is justified by the gain in constraining power with progressively larger dimensionality of the input phase-space distribution, as substantiated by smaller log-normal residual scatter and improved precision (cf. Figs.~\ref{fig:CNN_comparison_plot} and \ref{fig:CNN_precision_plot}, respectively). Compared to our recently proposed neural flow mass estimator \citep{DKR2020NF}, our CNN$_{\mathrm{3D}}$ model is more robust to the size of galaxy samples with spectroscopic redshifts, i.e. galaxy selection effects. The former method employs normalizing flows, implemented via a stack of multilayer perceptrons, to predict the posterior cluster mass PDFs from 2D phase-space distributions $\{ R_{\mathrm{proj}}, v_{\mathrm{los}} \}$, thereby deriving uncertainties in a conceptually distinct approach.

\medskip
The performance of our CNN$_{\mathrm{3D}}$ mass estimator, along with that of our recent neural flow mass estimator and the variational inference approach by \citet{ho2020approximate}, provides exciting avenues to infer cosmological constraints from the SDSS catalogue using cluster abundances \citep{abdullah2020cosmo}. These three novel machine learning algorithms yield robust and reliable cluster masses with complementary ways of deriving uncertainties and, therefore, may be utilized to constrain the cluster mass function to complement standard approaches based on traditional mass estimators.

\section*{Acknowledgements}

We express our gratitude to the reviewer whose detailed feedback helped to improve various aspects of our analysis. We acknowledge Tom Charnock for providing us with the motivation to make use of simulation-based inference with neural networks. We also thank Matthew Ho for his insights pertaining to the use of KDE for preprocessing the phase-space distribution and for his constructive feedback on our work. We are grateful to Guilhem Lavaux and Jens Jasche for interesting discussions related to Bayesian posterior validation. We also express our appreciation to Gary Mamon for his perceptive comments on our manuscript. DKR is a DARK fellow supported by a Semper Ardens grant from the Carlsberg Foundation (reference CF15-0384). This work was supported by a VILLUM FONDEN Investigator grant (project number 16599). This work has made use of the Horizon and Henon Clusters hosted by Institut d'Astrophysique de Paris. This work has been done within the activities of the Domaine d'Int\'er\^et Majeur (DIM) Astrophysique et Conditions d'Apparition de la Vie (ACAV), and received financial support from R\'egion Ile-de-France. This work made use of an HPC facility funded by a grant from VILLUM FONDEN (project number 16599).

\section*{Data availability}

The source code repository, containing \textsc{jupyter} tutorial notebooks and the mock cluster catalogue, is available at \url{https://github.com/doogesh/SBI_dynamical_mass_estimator}.




\bibliographystyle{mnras} 
\bibliography{./compiled_references} 



\appendix

\section{Cluster mass function}

\begin{figure}
	\centering
    \includegraphics[width=\hsize, clip=true]{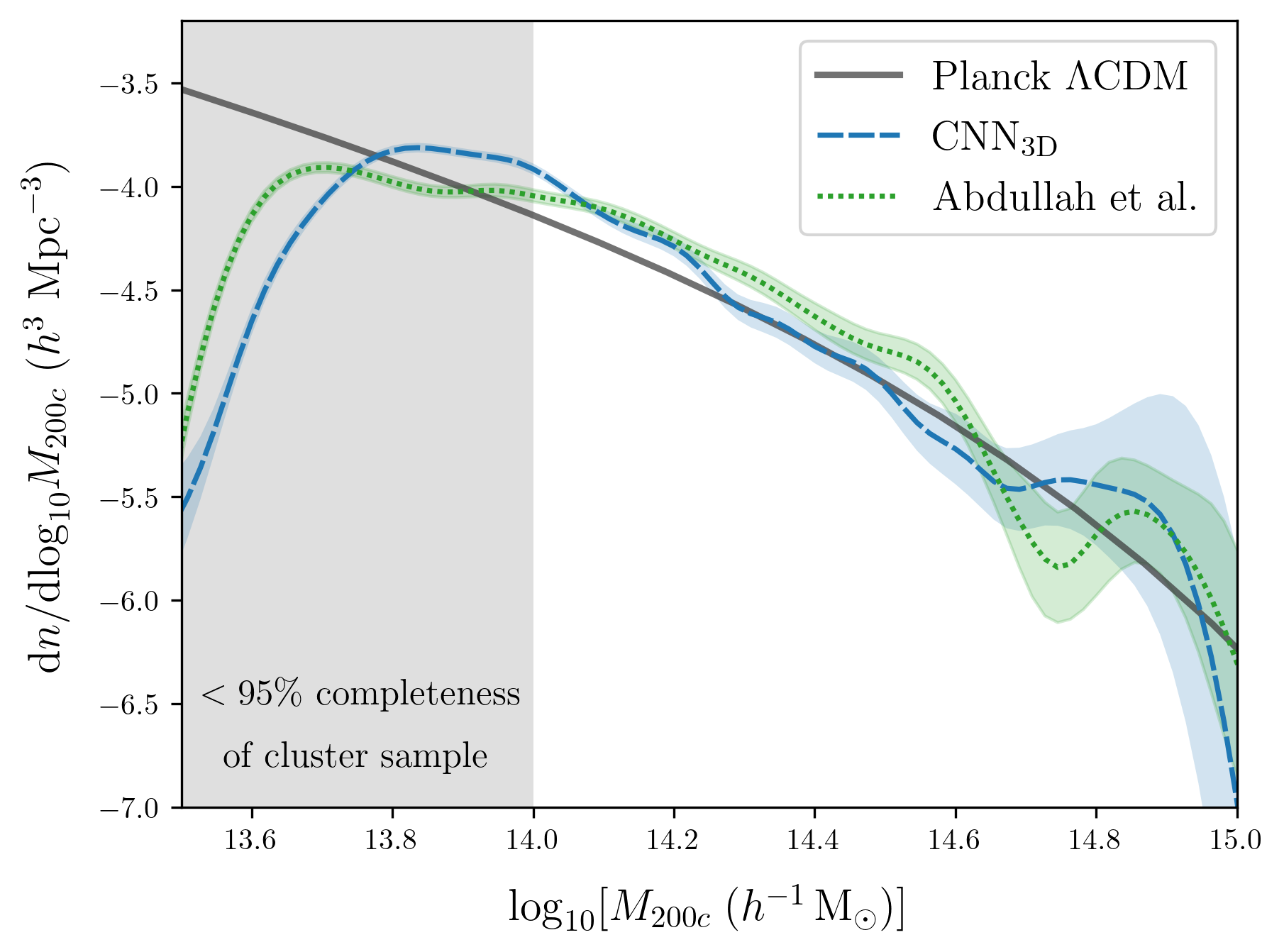}
	\caption{Cluster mass function derived from the dynamical mass measurements obtained for a sample of 760 SDSS galaxy clusters at redshifts $z\leq0.085$ using the new cluster mass estimation method devised in this work ($\rm{CNN}_{\rm{3D}}$). The result is compared to the corresponding cluster mass function computed for alternative mass estimates from \citet{Abdullah2020} based on the virial theorem and the theoretical halo mass function as predicted for Planck $\Lambda$CDM cosmology. The lines show the measured mass function obtained with a kernel density estimator, while the shaded bands indicate $1\sigma$ confidence interval from Poisson errors.
    The shaded gray region corresponds to the approximate mass range where the cluster sample is incomplete.}
	\label{fig:MF}
\end{figure}

This section constitutes non-peer reviewed supplementary material that is not included in the published version.

\medskip
Fig. \ref{fig:MF} shows the cluster mass function derived from our measurements of dynamical masses and those derived by \citet{Abdullah2020}. In both cases, we used the same sample of 760 galaxy clusters found in the SDSS footprint reduced by the perimeter area containing clusters affected by incompleteness due to proximity of the survey's edge. The total area of the reduced SDSS footprint is 6670 square degrees. Since the tests of the \textit{GalWeight} cluster finder based on both mock and real SDSS observations \citep{Abdullah2020,abdullah2020cosmo} ensure that the sample of rich clusters detected in the SDSS data is complete up to the maximum comoving distance adopted in our study ($z\leq0.085$), we compute the mass function assuming a constant selection function (weights equal to 1 for all clusters). The same tests also show that a minimum cluster mass for which the cluster finder is more than 95 per cent complete within the considered comoving volume is approximately $10^{14.0}h^{-1} {\rm M}_{\odot}$. We indicate the corresponding mass incompleteness range with a shaded area.

\medskip
The cluster mass functions computed from the two mass estimators are fully consistent. It is also readily apparent that both estimates of the cluster mass function recover the halo mass function of the Planck cosmological model \citep{planck2014cosmo} with a universal fitting function from \citet{Tinker2008} down to the approximate mass limit of the cluster sample completeness, i.e. $\sim10^{14.0}h^{-1} {\rm M}_{\odot}$.

\bsp	
\label{lastpage}
\end{document}